\documentclass[twocolumn,showpacs,preprintnumbers,amsmath,amssymb,pre]{revtex4}

\usepackage{graphicx}
\usepackage{dcolumn}
\usepackage{bm}

\begin{document}

\title{Elastic energy of polyhedral bilayer vesicles}

\author{Christoph A. Haselwandter and Rob Phillips}

\affiliation{Department of Applied Physics, California Institute of Technology, Pasadena, CA 91125, USA}

\date{\today}

\begin{abstract}

In recent experiments [M. Dubois, B. Dem\'e, T. Gulik-Krzywicki, J.-C. Dedieu,
C. Vautrin, S. D\'esert, E. Perez, and T. Zemb, Nature (London) \textbf{411}, 672 (2001)] the spontaneous formation of hollow bilayer vesicles with polyhedral
symmetry has been observed. On the basis of the experimental phenomenology
it was suggested [M. Dubois, V. Lizunov, A. Meister, T. Gulik-Krzywicki,
J. M. Verbavatz, E. Perez, J. Zimmerberg, and T. Zemb, Proc. Natl. Acad.
Sci. U.S.A. \textbf{101}, 15082 (2004)] that the mechanism for the formation of bilayer polyhedra is minimization of elastic bending energy. Motivated by these experiments, we study the elastic bending energy of polyhedral bilayer vesicles. In agreement with experiments,
and provided that excess amphiphiles exhibiting spontaneous curvature
are present in sufficient quantity, we find that polyhedral bilayer vesicles can
indeed be energetically favorable compared to spherical bilayer vesicles. Consistent with experimental observations we also find that the bending
energy associated with the vertices of bilayer polyhedra can be locally reduced through the formation of pores. However, the stabilization
of polyhedral bilayer vesicles over spherical bilayer vesicles relies crucially on molecular segregation of excess amphiphiles along the ridges rather than the vertices of bilayer polyhedra. Furthermore, our analysis
implies that, contrary to what has been suggested on the basis of experiments, the icosahedron does not minimize elastic bending energy among
arbitrary polyhedral shapes and sizes. Instead, we find that, for large polyhedron sizes, the snub dodecahedron and the snub cube both have lower total bending energies than the icosahedron.

\end{abstract}


\maketitle

\section{Introduction}

The self-assembly of complex two-dimensional objects from simple constituent units plays an important role throughout condensed matter physics, materials science, and molecular biology. Of particular importance for biology is the
self-assembly of amphiphilic molecules into flexible bilayers \cite{safran03,boal02,phillips09}
which provide the structural basis for cell membranes. The physical properties
of amphiphile bilayers are often studied using artificial bilayer vesicles
\cite{safran03,boal02,seifert97,phillips09}
of controlled molecular composition. In many settings
\cite{safran03,boal02,seifert97,wortis02,phillips09} the shape of such bilayer vesicles is characterized by a constant or smoothly varying curvature and minimizes the elastic energy of the vesicle. In experiment as well as theory \cite{boal02,seifert97,wortis02}, characteristic sequences of distinct vesicle shapes are obtained as a function of geometric parameters, such as the vesicle surface area at fixed vesicle volume, and elastic parameters, such as the bilayer spontaneous curvature. This has led to a general framework for the description and prediction of
smooth vesicle shapes \cite{boal02,seifert97} in which elasticity theory is combined with variational and perturbative methods for energy minimization.

In recent experiments \cite{dubois01,dubois04,meister03,glinel04,vautrin04}, however, facetted bilayer vesicles with shapes reminiscent of polyhedra have been observed.  Polyhedra are characterized by flat faces connected by ridges and vertices with high local curvature, and are generally not regarded as being energetically favorable shapes of bilayer vesicles. In these experiments, two types of oppositely charged, single-tailed amphiphiles were used \cite{dubois01,dubois04}, with a slight excess of one amphiphile species over the other. Consistent with
the classic view of bilayer vesicles \cite{safran03,boal02,seifert97,phillips09}, the amphiphiles were found to self-organize into spherical bilayer vesicles at high temperatures. However, provided that the number of excess, unpaired amphiphiles was tuned to some optimal range \cite{dubois01,dubois04,meister03,glinel04,vautrin04}, cooling the system below the chain melting temperature yielded the spontaneous formation of polyhedral bilayer vesicles. The bilayer polyhedra were reported to be stable over weeks, and to be consistently reproduced upon thermal cycling. Furthermore, it was suggested \cite{dubois01,dubois04} that the observed
polyhedral shapes had icosahedral symmetry, although some uncertainty regarding the polyhedral symmetry remained. Finally, the vertices of polyhedral bilayer vesicles were found to exhibit
pores \cite{dubois01,dubois04,glinel04}, which was put forward \cite{dubois01} as a mechanism for avoiding the large elastic bending energy associated with closed vertices of bilayer polyhedra.

On the basis of the experimental phenomenology it was suggested \cite{dubois01,dubois04} that minimization of elastic bending energy determines the shape of bilayer polyhedra. In a previous article \cite{haselwandter10} we took these intriguing experimental observations as our starting point and investigated the minimal bending energies of bilayer polyhedra. We found that, while polyhedral vesicles can be energetically favorable
compared to spherical vesicles for the bilayer composition used in the aforementioned experiments \cite{dubois01,dubois04,glinel04,meister03,vautrin04}, the snub dodecahedron and the snub cube generally have lower elastic bending energies than the icosahedron. The purpose of the present article is to provide a more comprehensive discussion of the bending energies of bilayer polyhedra for various experimental scenarios, and allowing for different models of the elastic contributions to the free energy of bilayer polyhedra. Our overall aim is thereby
to provide basic estimates of the relative bending energies associated with different polyhedral symmetries, and to contrast these polyhedral bending energies with the elastic bending energy of spherical bilayer vesicles. The
disagreement between experiment and theory concerning the most favorable polyhedral
symmetry suggests that either the mechanism governing the
shape of bilayer polyhedra is not solely minimization of elastic bending energy, or the
dominant shape of the facetted bilayer vesicles observed in experiments does
not correspond to the icosahedron.

To predict polyhedral shapes with minimal energy, a number of methodologies based on computer simulations have been developed over recent years \cite{dodgson96,bruinsma03,zandi04,vernizzi07,levand09}. Here we use a complementary method, in which we first derive general expressions
for the contributions to the elastic bending energy of bilayer polyhedra due to the ridges, closed vertices, and vertex pores observed experimentally \cite{dubois01,dubois04,glinel04}. Particularly simple expressions of ridge, vertex, and pore energies are obtained from the Helfrich-Canham-Evans free energy of bending \cite{canham70,helfrich73,evans74}. We assess the validity of these
phenomenological expressions, which only involve a few parameters, by making comparisons
to solutions of the two-dimensional equations of elasticity obtained previously
for the ridges and vertices of polyhedra in certain
limiting cases \cite{lobkovsky96,lobkovsky97,seung88,lidmar03,nguyen05,didonna02}.
On this basis we then survey total polyhedral bending energies for a variety of different symmetry classes of polyhedra \cite{cromwell97,coxeter80,weisstein09},
which are characterized by distinct values of the geometric parameters entering our expressions of ridge, vertex, and pore energies.

The organization of this article is as follows. Section~\ref{secFreeE} provides a brief review of the experimental phenomenology of bilayer polyhedra and of the contributions to their free energy. In Sec.~\ref{secBendE} we
derive general expressions for the elastic bending energies associated with
ridges, closed vertices, and vertex pores from the Helfrich-Canham-Evans free energy of bending. Comparisons to
the corresponding
solutions obtained previously in limiting cases of the equations of elasticity are made in Sec.~\ref{secAsym}. Section~\ref{secAnPore} analyzes the elastic bending energy associated with pores of bilayer polyhedra. In
Sec.~\ref{secPolyE} we calculate total bending energies of bilayer polyhedra
for various polyhedral symmetry classes. A discussion of our results
is provided in Sec.~\ref{secDisc}, and a summary and conclusions can be found
in Sec.~\ref{secSum}.

\section{Experimental phenomenology of bilayer polyhedra}
\label{secFreeE}

The bilayer polyhedra observed in experiments \cite{dubois01,dubois04,glinel04,meister03,vautrin04}
were composed of two different types of amphiphiles: myristic acid and cetyltrimethylammonium hydroxide (CTAOH). Myristic acid carries a single negative charge while CTAOH is positively charged, and the hydrophobic parts of both amphiphile species consist of a single hydrocarbon chain. In a salt-free aqueous solution dilute in amphiphiles, the two amphiphile
species were observed to self-assemble into bilayers \cite{zemb99}. The bilayers had a thickness of approximately 4~nm, and the inter-amphiphile
spacing was found \cite{zemb99,dubois01} to be around 0.4--0.6~nm. While above the chain melting temperature the bending rigidity of the bilayers formed by myristic acid and CTAOH falls within the range 1--10 $k_B T$, cooling the system to room
temperature yielded very stiff bilayers with rigidities greater than 100~$k_B T$ \cite{dubois04}. In small-angle neutron scattering experiments
it was indeed found \cite{zemb99} that bilayers were nearly flat over a spatial length scale of more than 1~$\mu$m.

In earlier work, a  mesoscopic model \cite{hartmann06} was used to further investigate the intriguing
mechanical properties of the catanionic bilayers summarized above. In this
model, electrostatic interactions are described by a standard Ising Hamiltonian,
while a spring network accounts for the formation of hydrogen bonds
between amphiphiles. The behavior of bilayers obtained with this model is consistent with a simple
picture \cite{zemb99,dubois01,dubois04} of catanionic bilayers in which oppositely
charged amphiphiles pair up to form zwitterionic amphiphiles with zero net charge and two hydrophobic tails, thereby expelling excess amphiphiles from flat bilayers. Due to their molecular shape, such unpaired excess amphiphiles
are expected to exhibit spontaneous curvature. It was estimated \cite{dubois04}
from the monolayer chain length of myristic acid
that the induced spontaneous curvature of excess anionic amphiphiles is equal
to around $0.3$~nm$^{-1}$.

At high temperatures, mixtures of myristic acid and CTAOH were found to self-assemble into spherical bilayer vesicles \cite{dubois04}. As the system is cooled below the chain-melting temperature, the behavior of these vesicles can be characterized \cite{zemb99,dubois01,dubois04,glinel04,meister03,vautrin04} by the fraction of the anionic amphiphile component over total amphiphile content, which we will denote by $r_I$. Using electron and light microscopy it was found that, if $r_I \neq 0.5$, spherical bilayer vesicles may facet to form polyhedral shapes or break up to form flat bilayer disks. Both of these aggregate shapes may coexist with spherical bilayer vesicles. While
the diameter of bilayer disks was observed \cite{zemb99} to vary from 30~nm to 3~$\mu$m,
the diameters of bilayer polyhedra were reported 
\cite{dubois01,dubois04,glinel04,meister03,vautrin04}
to fall within a characteristic
range of 1--2~$\mu$m. Bilayer polyhedra are estimated
\cite{dubois01} to contain around $10^7$ catanionic pairs and an excess of myristic acid corresponding to around $10^6$ single amphiphiles.

Electron and fluorescence microscopy studies have suggested \cite{dubois01,dubois04,glinel04,meister03,vautrin04} that bilayer polyhedra exhibit pores at their vertices. By bleaching fluorescent molecules inside bilayer polyhedra and measuring fluorescence recovery after photobleaching, the pore diameter was estimated \cite{glinel04} to be equal to around 40~nm. However, in the same set of experiments it was also found that some of the observed polyhedral
vertices were in fact closed. Finally, on the basis of electron and confocal
microscopy a classification of
the symmetry of bilayer polyhedra was attempted. Some features
of the observed polyhedral shapes, including their hexagonal cross section and five-fold vertex geometry, were found to be consistent with an icosahedral symmetry, but there was also considerable heterogeneity in the observed polyhedral shapes \cite{dubois01,dubois04,glinel04,meister03,vautrin04}.

Following Ref.~\cite{dubois04}, we distinguish between three basic types
of contributions to the free energy of bilayer polyhedra. First, there are elastic contributions to the free energy associated with the energy required
to bend
amphiphile bilayers along the ridges and closed vertices of polyhedra, and to bend
amphiphile monolayers along the edges of polyhedral pores. For a given polyhedral symmetry and size, the total energetic
cost associated with these terms depends
on the elastic parameters characterizing
the bilayer and on the geometric parameters defining the polyhedral
shape. The total elastic energy of bilayer polyhedra is to be compared with the elastic energy associated with bilayer vesicles exhibiting
a constant or smoothly varying curvature. In particular, if no constraints
on the vesicle surface area or the vesicle volume are imposed, and the bilayer composition
is homogeneous, the classic framework for the description of smooth vesicle shapes implies \cite{seifert97}
that spherical bilayer vesicles minimize bending energy.

A second class of contributions to the free energy of bilayer polyhedra arises from the entropic cost of segregating excess amphiphiles along polyhedral ridges and vertices. These entropic terms make the formation of bilayer polyhedra with defined amphiphile domains unfavorable compared to homogeneous bilayer vesicles. However, considering the experimental observation \cite{dubois04} of segregated domains of excess amphiphiles along polyhedral ridges and vertices, entropic contributions do not seem to be dominant. Indeed, the picture of heterogeneous bilayers presented in Refs.~\cite{zemb99,dubois01,dubois04,hartmann06} suggests that excess amphiphiles are segregated during the cooling down process, leading to the separation of amphiphile bilayers into distinct domains which do not mix at low temperatures. Here, we will not be concerned with the precise mechanism leading to the segregation
of amphiphile domains, and assume that bilayers formed by myristic acid and CTAOH do indeed spontaneously expel excess amphiphiles during the cooling down process.

Third, we need to consider electrostatic contributions to the free energy
of bilayer polyhedra. On the one hand, segregated excess amphiphiles carry charges of equal sign and, hence, repel each other. Thus, in addition to entropic effects, the mechanism leading to amphiphile segregation must overcome
electrostatic repulsion between excess amphiphiles. On the other hand, the finite surface charge
density observed along the ridges and vertices of bilayer polyhedra \cite{dubois04} induces
screening clouds in the surrounding solution. Different membrane shapes
lead to different shapes of the screening clouds which, as discussed further in Sec.~\ref{secGen}, can affect
the energetic cost associated with polyhedral ridges and pores. However,
electrostatic contributions to the bending rigidity of amphiphilic bilayers
are expected \cite{winter88,pincus90,winter92,brotons05} to be of the order of 1--10 $k_B T$ and, hence, at
least one order of magnitude smaller than the experimental values \cite{dubois04} of the bending
rigidity of bilayer polyhedra. This suggests \cite{hartmann06} that the electrostatic energies associated
with deformations of the screening cloud are small compared to the membrane
bending energy. Thus, we will follow here Ref.~\cite{dubois04}
and assume that the shape of bilayer polyhedra is governed by minimization
of elastic bending energy.

\section{Bending energies of bilayer polyhedra}
\label{secBendE}

In this section we derive simple phenomenological expressions for the bending
energies associated with the ridges, closed vertices, and vertex pores of bilayer polyhedra. Our starting point is  the Helfrich-Canham-Evans
free energy of bending \cite{boal02,phillips09,seifert97,safran03,canham70,helfrich73,evans74}, namely, 
\begin{equation} \label{helfrich}
G=\frac{K_b}{2} \int dS \left(\frac{1}{R_1}+\frac{1}{R_2}-H_0\right)^2\,,
\end{equation}
where $K_b$ is the bilayer bending rigidity, $R_1$ and $R_2$ are the two principal radii of curvature, and $H_0$ is the bilayer spontaneous curvature. In situations where we consider amphiphile monolayers instead of amphiphile bilayers, $K_b$ in Eq.~(\ref{helfrich}) is replaced by
the monolayer bending rigidity $K_b^\star$, and $H_0$ is replaced by
the monolayer spontaneous curvature $H_0^\star$.

\subsection{Ridge energy}
\label{secBendR}

\begin{figure}[!]
\center
\includegraphics[width=5.3cm]{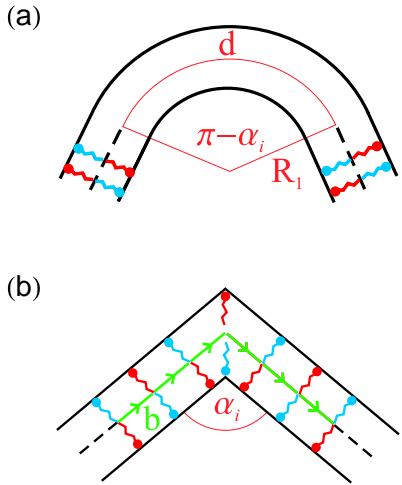}
 \caption{\label{FigRidgeE}(Color online) Side view of a ridge with dihedral angle $\alpha_i$ bending (a) over an arc length $d$ of a cylinder with radius $R_1$ and (b) over an arc length comparable to the small-scale cutoff $b$.
The red and blue amphiphile species represent myristic acid and CTAOH, which are negatively and positively charged, respectively. The arrows in panel (b) denote bond vectors connecting adjacent amphiphiles.}
\end{figure}

Figures~\ref{FigRidgeE}(a,b) show two models of an amphiphile bilayer bending along the ridge of a polyhedron with dihedral angle $\alpha_i$. For simplicity
we take $R_2 \to \infty$ in Eq.~(\ref{helfrich}) for both models. The elastic
energy of ridges
which do not necessarily satisfy this assumption will be discussed in Sec.~\ref{secAsym}.
Moreover, we focus here on the most straightforward case of symmetric
bilayer leaflets and take $H_0=0$~nm$^{-1}$ in Eq.~(\ref{helfrich}). The richer
case in which there is segregation of excess amphiphiles and, hence, the
possibility of an inhomogeneous composition of the membrane leaflets,
will be considered in Sec.~\ref{secSeg}.

Our first model in Fig.~\ref{FigRidgeE}(a) is inspired by the electron micrographs of bilayer polyhedra in Refs.~\cite{dubois01,dubois04,glinel04}. We assume that, along a ridge, a bilayer bends over an angle $\pi-\alpha_i$ around a cylinder of radius $R_1$, where the index $i$ denotes the particular polyhedral ridge under
consideration. From Eq.~(\ref{helfrich}) one then finds the ridge energy
\begin{equation} \label{ridgeE1}
G_{r}=\frac{K_b}{2} \frac{l_i}{d} \left(\pi-\alpha_i\right)^2\,,
\end{equation}
where $l_i$ is the ridge length and $d=R_1 \left(\pi-\alpha_i\right)$ is the arc length subtended by the ridge.

Our second model, illustrated in Fig.~\ref{FigRidgeE}(b), allows bilayers
to bend sharply along ridges and thereby provides a more faithful representation of the polyhedral geometry. We discretize the system \cite{kawakatsu04} using an inter-amphiphile spacing $b$ and note that, for a curve embedded
in two-dimensional space, the bond vector connecting adjacent amphiphiles is $\hat{\mathbf{t}}=(\cos
\phi, \sin \phi)$, where $\phi=\phi(u)$ is the angle between $\hat{\mathbf{t}}$
and the abscissa at some segment $u$ along the curve. Using the relation
\begin{equation} \label{ridgeProv2}
\frac{1}{R_1} = \frac{1}{b} \left| \frac{d\hat{\mathbf{t}}}{du} \right|
\end{equation}
one then obtains from Eq.~(\ref{helfrich}) a simple expression of the ridge
energy,
\begin{equation} \label{eqintstep1}
G_r=\frac{K_b}{2 b^2} \int dl \int d(b u) \left(\frac{d \phi}{d u} \right)^2\,.
\end{equation}
If the ridge bends ``sharply,'' we take
\begin{equation} \label{eqintstep2}
\left(\frac{d \phi}{d u}\right)^2=(\pi-\alpha_i)^2 \delta(2u)\,,
\end{equation}
where the factor of two in the argument of the Dirac delta function arises because we assume that the ridge in Fig.~\ref{FigRidgeE}(b)
bends over a length $2b$ so that a single amphiphile is located at the
tip of the ridge, thereby reducing the density in Eq.~(\ref{eqintstep1}). The ridge energy in Eq.~(\ref{ridgeProv2}) then becomes
\begin{equation}  \label{ridgeE2}
G_r=\frac{K_b}{4} (\pi-\alpha_i)^2 \frac{l_i}{b} \,.
\end{equation}
Setting $d=2b$, Eqs.~(\ref{ridgeE1}) and~(\ref{ridgeE2}) both yield
\begin{equation}  \label{ridgeE}
G_r^{(h)}=\frac{\bar K_b}{2} (\pi-\alpha_i)^2 l_i\,,
\end{equation} 
where the rescaled bilayer bending rigidity $\bar K_b=K_b/(2b)$ and the superscript $(h)$ indicates that this expression of the ridge energy applies to homogeneous membranes. An expression similar \cite{duboisTypo} to Eq.~(\ref{ridgeE}) was used in Ref.~\cite{dubois04} to describe the elastic bending energy associated with polyhedral ridges. The scale of the ridge energy in Eq.~(\ref{ridgeE}) is set by our assumption $d=2b$ which, on the basis of the scattering measurements in Refs.~\cite{dubois01,zemb99}, gives $d\approx1$~nm. The choice $d=2b\approx1$~nm
for the arc length,
and the resulting estimates of the energy density, are confirmed in Sec.~\ref{secAsym} by comparing Eq.~(\ref{ridgeE}) to an
expression of the ridge energy which allows for both principal radii of
curvature to be finite.

\subsection{Vertex energy}
\label{SecVertexE}

\begin{figure}[!]
\center
\includegraphics[width=7cm]{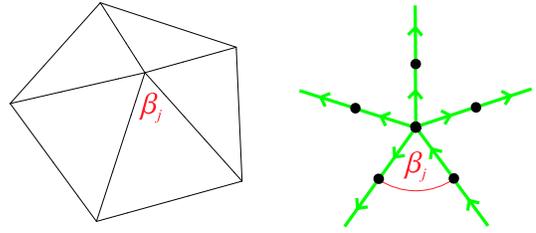}
 \caption{\label{FigRidgeV}(Color online) Illustration of a polyhedral vertex
with face angle $\beta_j$. As in Fig.~\ref{FigRidgeE}(b), the arrows represent
bond vectors connecting adjacent amphiphiles, but now with the bond vectors being
parallel rather than perpendicular to ridges.}
\end{figure}

A phenomenological but straightforward expression for the elastic bending
energy associated with closed bilayer vertices is obtained
following similar steps as in Sec.~\ref{secBendR}. As indicated in Fig.~\ref{FigRidgeV},
one can regard vertices as points at which the bond vectors parallel to ridges change direction to become parallel to neighboring ridges, which is complementary
to the model of ridges illustrated in Fig.~\ref{FigRidgeE}(b). We decompose the total vertex energy $G_v$ into a sum of $q$ terms $G_v^{(j)}$ with $j=1,\dots,q$, where $q$ denotes the number of ridges meeting at a vertex. Retracing the
steps leading to the ridge energy in Eq.~(\ref{ridgeE}) one finds
\begin{equation}
G_v^{(j)}=\frac{K_b}{2 b^2} \int dS \left( \frac{d{\phi}}{du} \right)^2=\frac{K_b}{2}
\left( \pi- \beta_j \right)^{2}\,,
\end{equation}
where $\beta_j$ denotes the face angle subtended at a given vertex by two neighboring ridges, and, similarly as in Sec.~\ref{secBendR}, we took the ridge length across a vertex to be equal to $2b$ and set
\begin{equation}
\left(\frac{d\phi}{du} \right)^2=\left( \pi- \beta_j \right)^{2} \delta\left(2u\right) \,.
\end{equation}
The total vertex energy is then given by
\begin{equation} \label{EVgen}
G_v^{(h)}=\sum_{j=1}^q G_v^{(j)}\,,
\end{equation} 
where the superscript $(h)$ again indicates that this expression applies to
homogeneous membranes.

\subsection{Pore energy}
\label{secPoreE}

\begin{figure}[!]
\center
\includegraphics[width=8cm]{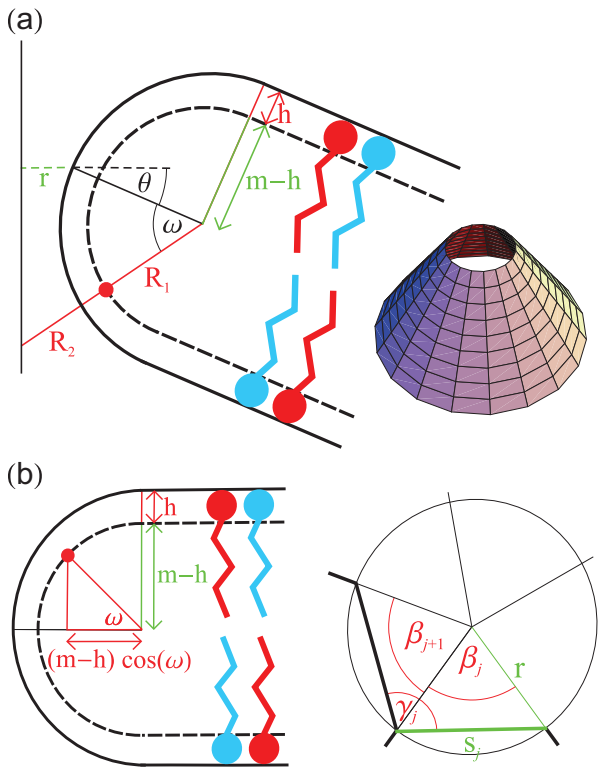}
 \caption{\label{figVertexP}(Color online) Schematic illustrations of two
models of vertex pores for an amphiphile monolayer thickness $m$ and an amphiphile
headgroup thickness $h$. (a) Cross section of half of a pore around the tip
of a cone (inset) with apex angle $\pi-2 \theta$ and radius $r$. (b) Side view (left panel) and top-down view (right panel) of a pore with radius
$r$ composed
of straight edges of length $s_j$ along each face which bend through an angle $\gamma_j$ from one face to a neighboring face.}
\end{figure}

Calculations of the elastic bending energy associated with toroidal pores in planar bilayers can be found in Refs.~\cite{chizmadzhev95,dubois04,glinel04}, and a generalization to arbitrary pore shapes is provided in Ref.~\cite{jackson09}.
Based on this previous work, we devised two complementary models of vertex pores. Our first model [see Fig.~\ref{figVertexP}(a)] is again inspired by the experimental images in Refs.~\cite{dubois01,dubois04,glinel04} which
suggest that the vertices of bilayer polyhedra locally resemble cones. Accordingly,
we approximate the vertex of a given polyhedron by a cone with apex angle $\pi-2\theta$, where $\theta=\pi/2-\arccos \left(1-\Omega/2 \pi \right)$ for a solid angle $\Omega$ subtended by the polyhedron vertex. We then use
the Helfrich-Canham-Evans free energy of bending in Eq.~(\ref{helfrich}) with the \textit{monolayer} bending rigidity and spontaneous curvature to
calculate the bending energy of a pore around the tip of a cone, leading
to an approximate expression for the bending energy of polyhedral pores.

From Fig.~\ref{figVertexP}(a) we read off the principal radii of curvature, $R_1$ and $R_2$, and the area element, $dS$, of a conical pore with semi-circular cross section:
\begin{eqnarray} \label{R1eq}
R_1&=&m-h\,,\\ \label{R2eq}
R_2&=&-\left(\frac{r+m \cos \theta}{\cos(| \omega|\pm\theta)}-m+h\right)\,,\\
dS&=&2 \pi R_1 \left[-R_2 \cos(|\omega|\pm\theta)\right] d\omega \,, \label{dSeq}
\end{eqnarray}
for $0 \leq \omega \leq \pi/2$ and $-\pi/2 \leq \omega \leq 0$, respectively,
where $m$ denotes the monolayer thickness, $h$ the thickness of the amphiphile headgroup, and $r$ is the pore radius. Note from Fig.~\ref{figVertexP}(a) that only pore radii $r\geq r_m$, where $r_m=m (1-\cos \theta)$ is defined as the value of $r$ for which the separation between opposite sides of the pore is equal to zero, have physical significance. For bilayer polyhedra,
we have \cite{dubois04} the representative values $m \approx 2$~nm and $h \approx 0.5$~nm.

Following the steps outlined in Appendix~\ref{AppPore}, one finds that Eqs.~(\ref{R1eq})--(\ref{dSeq}),
together with Eq.~(\ref{helfrich}), give
\begin{eqnarray}
G_p^{(c)}(r,\theta)&=&\pi K_b^\star  \nonumber
  \left[W(\xi,\theta)+ W(\xi,-\theta)+2
\mathcal{T}(\xi,\theta,H_0)\right]\,,\\ &&
 \label{GpFcal}
\end{eqnarray}
where we have defined
\begin{equation}
\xi\equiv\frac{r+m \cos \theta}{m-h}>1
\end{equation}
and
\begin{widetext}
\begin{eqnarray}
W(\xi,\theta)&=&\frac{2 \xi^2}{\left(\xi^2-1\right)^{1/2}} \left( \arctan \frac{\left(\xi^2-1\right)^{1/2}
\tan \left(\frac{\theta}{2}+\frac{\pi}{4}\right)}{\xi-1}- \arctan \frac{\left(\xi^2-1\right)^{1/2}
\tan \frac{\theta}{2}}{\xi-1}\right)\,,
\\
\mathcal{T}(\xi,\theta,H_0^\star)&=&-4\cos \theta
- H_0^\star (m-h) \left( \pi\xi-4  \cos \theta\right)
 +{H_0^\star}^2 (m-h)^2 \left(\frac{\pi}{2}\xi- \cos \theta \right)\,.
\end{eqnarray}
\end{widetext}
The superscript $(c)$ in Eq.~(\ref{GpFcal}) indicates
that this expression of the pore energy applies to conical pores. For $\theta=0$, the above result for the bending energy of a conical pore
reduces to the corresponding expression obtained previously for planar bilayers \cite{chizmadzhev95,dubois04,glinel04}.

Our second model of vertex pores [see Fig.~\ref{figVertexP}(b)] allows
for
a more faithful representation of the polyhedral geometry. We assume that, along each
face, the vertex pore consists of a straight edge with a semi-circular cross section [see Fig.~\ref{figVertexP}(b), left panel], which bends through an angle $\gamma_j$ across a ridge from one face to a neighboring
face [see Fig.~\ref{figVertexP}(b), right panel].
Accordingly, we split the energy cost associated with such a polygonal pore
into a term $G_p^{(1)}$ corresponding to the elastic bending energy of a straight
edge along a polyhedral face, and a term $G_p^{(2)}$ corresponding to the
energy cost of bending the edge of the pore from one face to
a neighboring face. For a straight edge of length $s_j$, the version of Eq.~(\ref{helfrich})
appropriate for a monolayer gives
\begin{equation} \label{GP1cont}
G_p^{(1)}=\frac{K_b^\star}{2} \pi (m-h) s_j \left(\frac{1}{m-h}-H_0^\star\right)^2\,,
\end{equation}
where from Fig.~\ref{figVertexP}(b) we have that $s_j=2 r \sin \frac{\beta_j}{2}$.

The contribution $G_p^{(2)}$ stems from bending the pore about the vertical axis in the left panel of Fig.~\ref{figVertexP}(b) by some angle $\gamma_j$. According to the right panel of Fig.~\ref{figVertexP}(b) we have
\begin{equation} \label{defGamma}
\gamma_j=\frac{1}{2} \left(2 \pi-\beta_j-\beta_{j+1} \right)\,.
\end{equation}
Retracing the steps which led to the ridge energy in Eq.~(\ref{ridgeE}),
but now for the horizontal amphiphile component as indicated in the left panel of Fig.~\ref{figVertexP}(b), one finds
\begin{equation} \label{GP2cont}
G_p^{(2)}=\bar K_b^\star (m-h) \left(\pi-\gamma_j +\bar H_0^\star\right)^2\,,
\end{equation}
where the rescaled monolayer bending rigidity $ \bar K_b^\star=K_b^\star/(2b)$ and the dimensionless spontaneous curvature $\bar H_0^\star=2b H_0^\star$. The total pore energy is then given by
\begin{equation} \label{polyPore}
G_p^{(p)}=\sum_{j=1}^q \left(G_p^{(1)} + G_p^{(2)} \right)\,,
\end{equation}
where the superscript $(p)$ signifies that Eq.~(\ref{polyPore}) applies to
a polygonal pore. Our expressions of conical and polygonal pore energies
in Eqs.~(\ref{GpFcal}) and~(\ref{polyPore}) will be discussed further in Sec.~\ref{secAnPore}.

\subsection{Segregation of excess amphiphiles}
\label{secSeg}

The experimental phenomenology of polyhedral bilayer vesicles suggests \cite{dubois01,dubois04} that the two amphiphile species constituting bilayer polyhedra pair
up to form flat bilayers and thereby expel excess (unpaired) amphiphiles
from polyhedral faces. As already noted in Sec.~\ref{secFreeE}, segregated excess amphiphiles exhibit a spontaneous curvature $H_0^\star \approx 0.3$~nm$^{-1}$
\cite{dubois04}, thus favoring a curved membrane shape. It has indeed been observed \cite{dubois01,dubois04,glinel04}
that excess amphiphiles seed pores into bilayers and localize along the ridges of bilayer polyhedra. As far as the effect of excess amphiphiles on pore
energies is concerned, we therefore follow Ref.~\cite{dubois04} and assume that vertex pores are composed of excess amphiphiles, leading to a finite value of $H_0^\star$ in Eqs.~(\ref{GpFcal}) and~(\ref{polyPore}) if sufficiently
many excess amphiphiles are present.

\begin{figure}[!]
\center
\includegraphics[width=7cm]{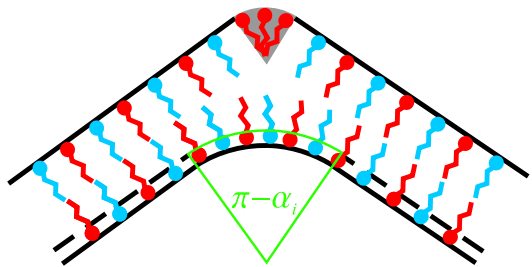}
 \caption{\label{FigRidgeESeg}(Color online) Side view of a ridge with dihedral angle $\alpha_i$ and perfect segregation of excess amphiphiles
in the outer bilayer leaflet. Note that, compared to Fig.~\ref{FigRidgeE}, the neutral
plane of bending is shifted from the mid-plane of the bilayer to the amphiphile
head-tail interface of the inner membrane leaflet.}
\end{figure}

How does segregation of excess amphiphiles modify the ridge energy in Eq.~(\ref{ridgeE})
and the vertex energy in Eq.~(\ref{EVgen})? To address this
question, we will consider a particularly favorable scenario for the formation of polyhedral ridges and vertices in heterogeneous
bilayers, and thereby obtain
a lower bound on the elastic energies associated with ridges and pores in
the presence of molecular segregation. Assuming
perfect segregation, excess amphiphiles are concentrated along the outer membrane
leaflets along ridges and closed vertices so as to induce an anisotropic spontaneous
curvature commensurate with the dihedral and vertex angles associated with
a given polyhedral geometry. As illustrated in Fig.~\ref{FigRidgeESeg} for
a ridge with dihedral angle $\alpha_i$, this leaves us with the inner membrane leaflet which, in the absence of some additional amphiphile species with
``inverted-wedge shape'' \cite{safran03,phillips09,boal02}, must be bent in order to cover the hydrophobic
tails of the excess amphiphiles localized in the outer membrane leaflet along
ridges and closed bilayer vertices.

For perfectly segregated ridges and vertices, we describe the bending of the inner membrane leaflet in a similar way as in the case
of the bilayer ridges and vertices considered in Secs.~\ref{secBendR} and~\ref{SecVertexE}, but with the neutral plane of bending shifted from the mid-plane of the bilayer to the amphiphile
head-tail interface of the inner amphiphile leaflet (see Fig.~\ref{FigRidgeESeg}). Thus, following analogous
steps as in Secs.~\ref{secBendR} and~\ref{SecVertexE}, we obtain the modified ridge and vertex energies
\begin{eqnarray} 
G_r^{(s)}&=&\frac{\bar K_b^\star}{2} (\pi-\alpha_i)^2 l_i\,, \label{modRE}\\
\label{modVE}
G_v^{(s)}&=&\frac{K_b^\star}{2} \sum_{j=1}^q \left( \pi- \beta_j \right)^{2}\,,
\end{eqnarray}
where the superscript $(s)$ indicates that in these expressions we assume perfect segregation of excess amphiphiles. Thus, provided that the optimal amount
of excess amphiphiles is present \cite{dubois01,dubois04}, the ridge and
vertex energies are lowered by a factor $K_b^\star/K_b$. Experiments \cite{dubois01,dubois04} and simulations \cite{hartmann06}
suggest that $K_b^\star/K_b\lessapprox10^{-2}$. Since the segregation of excess amphiphiles, and their fit to dihedral and vertex angles, will generally be less than perfect, we regard the simple phenomenological
expressions in Eqs.~(\ref{modRE}) and~(\ref{modVE}) as lower bounds on the ridge and vertex energies in heterogeneous bilayers.

The heuristic picture of amphiphile segregation developed above allows us to estimate the amount of excess amphiphiles present for a given polyhedral shape and size. In particular, we define the fraction of anionic amphiphile
content, which is the amphiphile species in excess for bilayer polyhedra
\cite{dubois01,dubois04,glinel04,meister03,vautrin04}, over total amphiphile content as
\begin{equation} \label{estrI}
r_I=\frac{1}{2}+\frac{N_R+N_P}{N_T}\,,
\end{equation}
where $N_R$ denotes the total number of excess amphiphiles segregated along ridges, $N_P$ denotes the total number of excess amphiphiles segregated
at vertex pores, $N_T$ denotes the total number of amphiphiles contained in the polyhedron shell, and, consistent with the estimates in Sec.~\ref{secFreeE}, we have taken $N_R+N_P\ll N_T$. In agreement with typical experimental observations \cite{dubois01,dubois04,glinel04}, Eq.~(\ref{estrI}) assumes that bilayer polyhedra exhibit pores at their vertices.

In order to estimate $N_R$ we need to determine how many excess amphiphiles must be segregated at a given polyhedral ridge so that the bilayer is bent by an appropriate dihedral angle. Assuming perfect segregation of excess
amphiphiles in the outer leaflet, we estimate that $n_i=(\pi-\alpha_i)/(b
H_0^\star)$ excess amphiphiles must be segregated per inter-amphiphile spacing
along the ridge in order to induce an angle $\pi-\alpha_i$ in the outer membrane leaflet. Thus,
\begin{equation} \label{estNS}
N_R=\sum_i l_i n_i\,,
\end{equation}
where the sum is to be taken over all the ridges of a given polyhedral shape.

A particularly simple estimate of $N_P$ is obtained by assuming that pores have a (flat) toroidal shape and radius $r$, which gives a pore surface area
of $2 \pi^2 m (m+r)$~nm$^2$. One therefore finds that
\begin{equation} \label{estNP}
N_P=V\frac{2 \pi^2 m (m+r)}{b^2}\,,
\end{equation}
where $V$ denotes the number of vertices of a given polyhedral shape and $b^2$ is the surface area per amphiphile. Similarly, a rough estimate of the total number of amphiphiles contained in the polyhedron shell is given
by
\begin{equation} \label{estNT}
N_T=\frac{8 \pi R_p^2}{b^2}\,,
\end{equation}
in which we have implicitly defined \cite{lidmar03} the polyhedron radius $R_p$ so that the polyhedron area is equal to $4 \pi R_p^2$ for a given edge length and polyhedral
symmetry. Combining Eqs.~(\ref{estNS})--(\ref{estNT}) we can evaluate the ideal value
of $r_I$ in Eq.~(\ref{estrI}) obtained from our simple description of amphiphile
segregation and make comparisons to the corresponding experimental estimates, a point we will return to in Sec.~\ref{secPolyE}.

\section{Asymptotic expressions of vertex and ridge energies}
\label{secAsym}

The solution of the two-dimensional equations of elasticity \cite{landau70}
is a formidable challenge, and has only been achieved for the ridges and vertices of polyhedra in certain limiting cases \cite{seung88,lidmar03,nguyen05,lobkovsky96,lobkovsky97,didonna02}
corresponding to a diverging F\"oppl-von K\'arm\'an number. The F\"oppl-von K\'arm\'an number is a dimensionless quantity characterizing the competition
between bending and stretching deformations and, for spherical shells, is defined as \cite{lidmar03}
\begin{equation}
\Gamma=\frac{Y R_p^2}{K_b}\,,
\end{equation}
where $Y$ is the two-dimensional Young's modulus. In the following we will use the available asymptotic solutions of the equations of elasticity for
polyhedral ridges and vertices to assess the validity of the phenomenological expressions of ridge and vertex energies obtained in Secs.~\ref{secBendR} and~\ref{SecVertexE}.

\subsection{Vertex energy}
\label{secVtable}

In a series of papers \cite{seung88,lidmar03,bowick00,nguyen05,bowick06}, the energetic cost of introducing five-fold disclinations in hexagonal lattices has been investigated. It was found \cite{seung88} that for a flat,
circular sheet of radius $R$, the stretching energy diverges linearly with the area of
the sheet: $E(R) = A_0 Y R^2$, where $A_0$ is a constant. However, if the sheet is
allowed to buckle out of the plane, it is, for large enough system sizes, energetically favorable to form
a cone with a central region which is ``flattened out,'' thus avoiding the
curvature singularity at the tip of the cone. The bending energy associated
with the cone section is found to be $E(R) = B_0 K_b \log
\left(R/R_b\right)$, where $B_0$ is a constant and $R_b$ is the buckling radius at which it becomes energetically favorable for the lattice to bend out of the plane.

The above results have been used to estimate \cite{lidmar03,nguyen05} the elastic energy of icosahedral vertices by noting that spherical shells can be discretized
using an icosadeltahedral triangulation, which consists of a hexagonal lattice
exhibiting twelve five-fold disclinations. Regarding the twelve disclination sites as  independent, the vertex energy of icosadeltahedral triangulations of the sphere is found to be given by
\begin{equation} \label{VestimateNelson}
\frac{E(\Gamma)}{K_b}\approx
\begin{cases} \frac{B_0}{2} \frac{\Gamma}{\Gamma_b} & \text{for $ \Gamma<\Gamma_b$,} \vspace*{0.1cm} \\ \frac{B_0}{2} \left[1+\log \left(\frac{\Gamma}{\Gamma_b} \right)\right] & \text{for
$ \Gamma>\Gamma_b$},
\end{cases}
\end{equation}
for each disclination \cite{lidmar03,nguyen05}, where $\Gamma_b \equiv Y R_b^2/K_b$ is the critical F\"oppl-von K\'arm\'an number
for buckling to occur, we have neglected constant contributions due to the
spherical
background curvature, and the parameter $A_0$ has been eliminated by
energy minimization with respect to $R_b$. Good fits \cite{lidmar03,nguyen05} to the results of simulations are obtained with $B_0 \approx 1.30$ and $\Gamma_b \approx 130$.

The energy in Eq.~(\ref{VestimateNelson}) corresponds, for large enough~$\Gamma$,
to the elastic energy associated with the vertex of the icosahedron and can, in this limit, be compared to the more general but heuristic vertex
energy in Eq.~(\ref{EVgen}) with values of the geometric parameters appropriate
for icosahedral vertices. To this end, we note that the lowest energy states of icosadeltahedral
triangulations of the sphere are found to resemble icosahedra for $\Gamma\gtrapprox10^7$
\cite{lidmar03,nguyen05}, which corresponds to a vertex energy greater than $8 K_b$ with, for instance, a value $12K_b$ for $\Gamma=10^{10}$. As shown
in Table~\ref{sumV}, this estimate compares quite favorably with the value $G_v \approx 11 K_b$ implied by Eq.~(\ref{EVgen}) for the icosahedron. In the estimates obtained from Eq.~(\ref{VestimateNelson}), the contribution due to stretching, which is not considered in Eq.~(\ref{EVgen}), is approximately
equal to $0.65 K_b$. Thus, the energetic cost associated with bending deformations
is seen to
dominate over the energetic cost associated with stretching deformations
in this regime of $\Gamma$.
For comparison, we note that for bilayer polyhedra it has been estimated \cite{dubois04} that $\Gamma \approx 10^6$, which, according to Eq.~(\ref{VestimateNelson}),
would leave us with a vertex energy of approximately $6.5 K_b$.

\begin{table}
\caption{\label{sumV} Vertex energy $G_v^{(h)}$ in Eq.~(\ref{EVgen}) for homogeneous bilayers, vertex energy $G_v^{(s)}$ in Eq.~(\ref{modVE}) for perfectly segregated bilayers, conical pore energy $G_p^{(c)}$ in Eq.~(\ref{GpFcal}) for the minimum
pore radius $r=r_m$, and polygonal pore energy $G_p^{(p)}$ in Eq.~(\ref{polyPore})
for $r=0$ in units of monolayer bending modulus $K_b^\star$ for the five Platonic solids with $m=2$~nm and $h=0.5$~nm \cite{dubois04}. The ranges
of $G_p^{(c)}$ and $G_p^{(p)}$
are obtained with $H_0^\star=0$~nm$^{-1}$ and $H_0^\star=0.3$~nm$^{-1}$, respectively.}
\begin{ruledtabular}
\begin{tabular}{lccccc}
Platonic solid  & $G_v^{(h)}$ & $G_v^{(s)}$ & $G_p^{(c)}(r=r_m)$ & $G_p^{(p)}(r=0)$ \\
\hline \vspace*{0.1cm}\\[-0.4cm]
Tetrahedron & $6.6\frac{K_b}{K_b^\star}$ & $6.6 $ & $10$--$15$ &  $4.9$--$8.2$ \vspace*{0.1cm}\\
Cube & $3.7\frac{K_b}{K_b^\star}$ & $3.7 $ & $8.9$--$11$
& $11$--$16 $ \vspace*{0.1cm}\\
Octahedron & $8.8\frac{K_b}{K_b^\star}$ & $8.8$ & $9.1$--$12$ & $6.6$--$11$ \vspace*{0.1cm}\\
Dodecahedron & $2.4\frac{K_b}{K_b^\star}$ & $2.4$ & $7.6$--$8.3$ & $16$--$21$ \vspace*{0.1cm}\\
Icosahedron & $11\frac{K_b}{K_b^\star}$ & $11$ & $7.9$--$8.9$
& $8.2$--$14$
\end{tabular}
\end{ruledtabular}
\end{table}

Note from Table~\ref{sumV} that the bending energies associated with (closed) conical
and polygonal pores take similar values for all Platonic solids, with the
competition between pores and closed, homogeneous bilayer vertices governed
by the ratio $K_b/K_b^\star$. Using the estimate $K_b/K_b^\star \gtrapprox10^2$ suggested by experiments \cite{dubois01,dubois04} and simulations \cite{hartmann06}, we find that closed bilayer vertices will be unstable to the formation of pores. However, Table~\ref{sumV} also implies that the vertex energy obtained
for perfectly segregated
bilayers is comparable to the bending energy associated with pores,
suggesting that, if the optimal amount of excess amphiphiles is present at
polyhedral vertices,
closed bilayer vertices may be metastable.

\subsection{Ridge energy}

According to computer simulations \cite{lidmar03,nguyen05}, the total energy
of icosadeltahedral triangulations of the sphere
is dominated by vertex energies for $\Gamma\lessapprox10^7$, and only in the regime of very large $\Gamma$, where the overall shape becomes
increasingly icosahedral, do the contributions of ridges to the overall elastic
energy become significant. It was shown by Lobkovsky and Witten \cite{lobkovsky96,lobkovsky97,lidmar03,nguyen05}
that, for $\Gamma \to \infty$, the ridge energy is given by 
\begin{equation} 
G_r^{(LW)} \approx 1.24 K_b \left(\frac{\pi-\alpha_i}{2}\right)^{7/3} \left(\frac{Y l_i^2}{K_b}\right)^{1/6}\,.
\end{equation}
Allowing the broad ranges $10^2 ~k_B T \lessapprox K_b \lessapprox10^4 ~k_B T$ and $10 ~k_B T/\mbox{nm}^2\lessapprox Y \lessapprox 10^3 ~k_B T/\mbox{nm}^2$
for bilayer polyhedra \cite{footnoteEr},
this implies 
\begin{equation} \label{ridgeELW}
0.1 K_b (\pi-\alpha_i)^{7/3} l_i^{1/3} \lessapprox G_r^{(LW)}\lessapprox 0.4 K_b (\pi-\alpha_i)^{7/3} l_i^{1/3}\,,
\end{equation}
which should be compared to Eq.~(\ref{ridgeE}). The ridge energies in Eqs.~(\ref{ridgeE}) and~(\ref{ridgeELW}) yield similar results for a unit ridge length, thus confirming the assumption $d=2b$ made in Sec.~\ref{secBendR}.
However, the estimate
of the ridge energy due to Lobkovsky and Witten has a stronger dependence
on the dihedral angle, but increases more slowly with $R_p$. In addition to the
ridge energy in Eq.~(\ref{ridgeE}), we will therefore also consider Eq.~(\ref{ridgeELW})
when calculating the total elastic energy of bilayer polyhedra in Sec.~\ref{secPolyE}.

\section{Analysis of pore energies}
\label{secAnPore}

In Sec.~\ref{secPoreE} we obtained Eqs.~(\ref{GpFcal}) and~(\ref{polyPore})
as the elastic bending energies associated with conical and polygonal pores.
The purpose of the present section is to discuss how these pore energies
vary with the elastic and geometric parameters characterizing polyhedral bilayer vesicles.

\subsection{Conical pores}
\label{secConPore}

\begin{figure}
\center
\includegraphics[width=8.5cm]{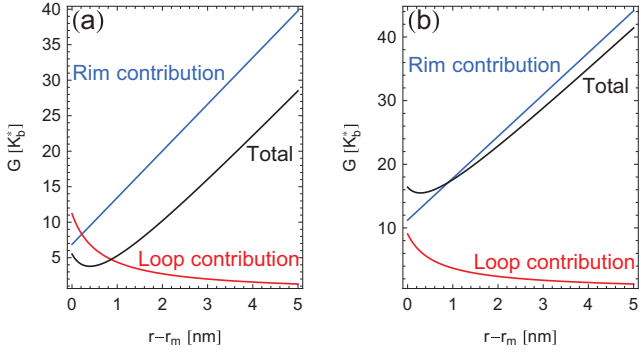}
\caption{\label{FigPoreContr}(Color online) Total bending energy of a conical
pore in Eq.~(\ref{GpFcal}), rim contribution in Eq.~(\ref{rimContrib}), and loop contribution in Eq.~(\ref{poreContrib}) versus pore radius
$r$ for \cite{dubois04} $m=2$~nm, $h=0.5$~nm, and $H_0^\star=0$~nm$^{-1}$, with (a) $\theta=0$ and (b) $\theta=0.4 \pi$.}
\end{figure}

Figures~\ref{FigPoreContr}(a,b) show plots of the conical pore energy in Eq.~(\ref{GpFcal})
as a function of the pore radius for $\theta=0$ and $\theta=0.4 \pi$, respectively. The angle $\theta=0$ corresponds to a
flat bilayer, while $\theta=0.4 \pi$ roughly corresponds to the vertex geometry
of the tetrahedron. A notable feature of the curves in Fig.~\ref{FigPoreContr} is that $G_p^{(c)}$ exhibits a minimum as a function of $r$. This optimal pore radius arises due to the competition between the standard edge tension
of a straight bilayer edge \cite{boal02} acting along the rim of the pore, which leads to an energy cost increasing linearly with~$r$, and the bending energy associated with closing the pore, which is expected
to be large for small pore radii. We can see this more clearly by returning to the Helfrich-Canham-Evans free energy of bending in Eq.~(\ref{helfrich}).
For simplicity we set $H_0^\star=0$~nm$^{-1}$, in which case the ``rim contribution'' corresponds to the integral over $1/R_1$ and
is given by
\begin{eqnarray} \nonumber
\mathcal{R}&=&- \pi K_b^\star \int_{-\pi/2}^{\pi/2} d \omega \cos(|\omega|\mp\theta)
\frac{R_2}{R_1}
\\&=& \pi K_b^\star \left(\pi \xi -2 \cos \theta\right)\,, \label{rimContrib}
\end{eqnarray}  
with $\xi \propto r$, where the principal radii of curvature $R_1$ and $R_2$
are defined in Eqs.~(\ref{R1eq}) and~(\ref{R2eq}), respectively. Similarly, the ``loop contribution'' is associated with the integral
over $1/R_2$ and evaluates to
\begin{eqnarray} \nonumber
\mathcal{L}&=&- \pi K_b^\star \int_{-\pi/2}^{\pi/2} d \omega \cos(|\omega
|\mp\theta)
\frac{R_1}{R_2}\\
&=&  \pi K_b^\star \left[-\pi \xi-2 \cos \theta+W(\xi,-\theta)+W(\xi,\theta)\right]\,.
\nonumber \\&& \label{poreContrib}
\end{eqnarray} 
Neglecting terms which are constant in $r$, the sum of $\mathcal{R}$ and $\mathcal{L}$
is equal to $G_p^{(c)}$ in Eq.~(\ref{GpFcal}) for $H_0^\star=0$~nm$^{-1}$.

Figure~\ref{FigPoreContr}(a) shows that, in the case $\theta=0$, the rim and pore contributions to $G_p^{(c)}$ do indeed behave as expected, with $\mathcal{R}$ increasing linearly with the pore circumference $2 \pi r$, and $\mathcal{L}$ decreasing with increasing $r$. The sum of $\mathcal{R}$ and $\mathcal{L}$ exhibits a minimum as a function of $r$. As illustrated in Fig.~\ref{FigPoreContr}(b), these characteristic features persist for $\theta>0$, with the small $r$ regime ($0<r\lessapprox2$~nm) dominated by the nonlinear behavior of the loop contribution to the bending energy, and the large $r$ regime ($r\gtrapprox2$~nm) dominated by the linear
behavior of the rim contribution. Considering that the optimal pore radius typically found from Eq.~(\ref{GpFcal}) is of the order of one nanometer, which is close to the smallest length scales down to which a description of bilayer pores in terms of continuum elasticity theory can be expected to apply~\cite{wohlert06}, it is questionable \cite{footnoteDGerror} whether the optimal pore radius exhibited by conical pores is of physical significance.

\begin{figure}
\center
\includegraphics[width=8.63cm]{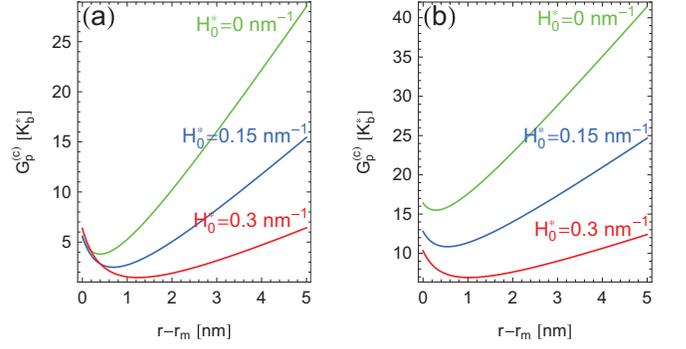}
\caption{\label{FigPoreEcone}(Color online) Bending energy of a conical pore in Eq.~(\ref{GpFcal}) versus pore radius $r$ for \cite{dubois04} $m=2$~nm, $h=0.5$~nm, and (a) $\theta=0$ and (b) $\theta=0.4 \pi$ for the indicated values of the monolayer spontaneous curvature
$H_0^\star$.}
\end{figure}

Figure~\ref{FigPoreEcone} compares the pore energies obtained with $\theta=0$ and $\theta>0$ for a number of different values of $H_0^\star$. For the range of spontaneous curvatures considered \cite{dubois04}, the pore
energy is seen to decrease with increasing monolayer spontaneous curvature. However,
for values of the spontaneous curvature much larger ($H_0^\star \gtrapprox 0.8$~nm) than those in Fig.~\ref{FigPoreEcone}, the rim curvature no longer
suffices to relax the spontaneous curvature and the pore energy rises again
with increasing $H_0^\star$. Moreover, from Figs.~\ref{FigPoreContr} and~\ref{FigPoreEcone} we observe the general trend that the conical pore energy is increased relative to the pore energy of
a planar bilayer, by up to approximately $13$ $K_b^\star$ for the parameter values used
in Figs.~\ref{FigPoreContr} and~\ref{FigPoreEcone}, with a larger increase in the pore energy corresponding to a smaller apex angle.

What is the physical origin of the increase in $G_p^{(c)}$ for $\theta>0$? Comparing panels (a) and (b) in Figs.~\ref{FigPoreContr} and~\ref{FigPoreEcone}
we note that the fractional difference between the pore energies for $\theta>0$
and $\theta=0$ is large for small pore radii, but decreases as the pore radius
increases. Indeed, the ratio $G_p^{(c)}(\theta>0,r)/G_p^{(c)}(\theta=0,r)$ approaches one as $r$ tends to infinity. In order to understand this
behavior on a qualitative level, note that $r_m=0$ for $\theta=0$, but $r_m>0$ for $\theta>0$.
In the latter case, if $r=r_m$, the inner sections of the pore ``almost'' touch,
and the loop contribution to the bending energy is large. For $\theta=0$, however, there
is still a pore of finite diameter at $r=r_m$, leading to a correspondingly
smaller loop contribution to the bending energy. As
$r$ becomes large, $G_p^{(c)}(\theta>0,r)$ and $G_p^{(c)}(0,r)$ are both increasingly dominated
by the rim contribution to the bending energy, and, thus, their ratio as
a function of the pore radius approaches one.

\begin{figure}
\center
\includegraphics[width=8.5cm]{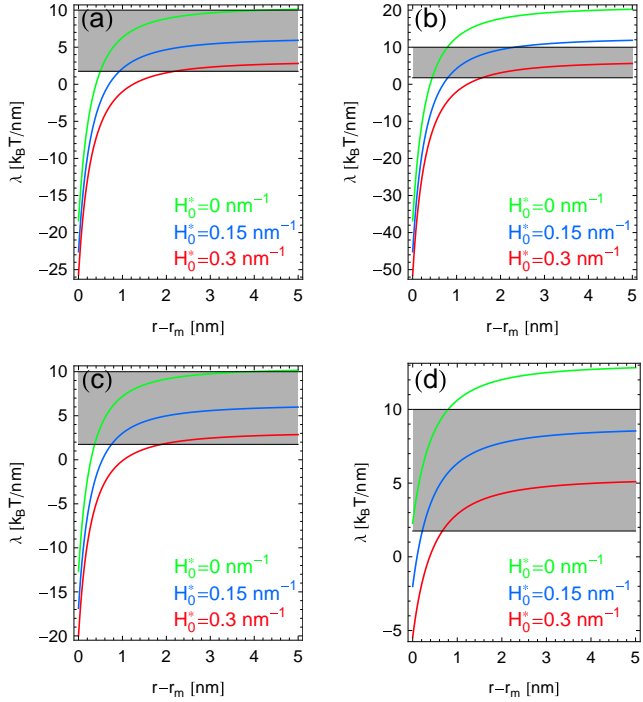}
\caption{\label{FigLineTensionC}(Color online) Edge tension $\lambda$ in Eq.~(\ref{ExprLTC}) for a conical pore versus pore radius $r$ for $m=2$~nm and (a) $\theta=0$ and $h=0.5$~nm with $K_b^\star=10k_B T$, (b) $\theta=0$ and $h=0.5$~nm with $K_b^\star=20k_B T$, (c) $\theta=0.4 \pi$ and $h=0.5$~nm with $K_b^\star=10k_B T$, and (d) $\theta=0$ and $h=0.8$~nm with $K_b^\star=10k_B T$ using $H_0^\star=0$~nm$^{-1}$ for the green (upper) curves, $H_0^\star=0.15$~nm$^{-1}$
for the blue (middle) curves, and $H_0^\star=0.3$~nm$^{-1}$ for the red (lower) curves in each panel. The shaded regions of the plots correspond to typical measured values \cite{boal02}
of the edge tension of amphiphile bilayers.}
\end{figure}

Finally, we use our expression of the pore energy in Eq.~(\ref{GpFcal})
to evaluate the edge tension, $\lambda$, associated with a conical pore:
\begin{equation} \label{ExprLTC}
\lambda\equiv \frac{\partial G_p^{(c)} }{\partial (2 \pi r)}=\frac{1}{2 \pi (m-h)}
\frac{\partial G_p^{(c)}}{\partial \xi}\,.
\end{equation}
Figure~\ref{FigLineTensionC} shows plots of the edge tension in Eq.~(\ref{ExprLTC})
as a function of the pore radius
and compares the calculated estimates to representative values of $\lambda$ measured in experiments \cite{boal02}. Experimental estimates of the edge
tension typically rely \cite{boal02} on measurement of the maximum disk size formed by lipid
bilayers, or on measurement of the pore radius in bilayers under tension. Our theoretical estimates of the edge tension vary depending on the choice for the numerical values of
$H_0^\star$ and $K_b^\star$, but are found to be in broad agreement with experimental measurements. A notable discrepancy between theoretical and experimental results is that the edge tension in Eq.~(\ref{ExprLTC}) depends
on $r$ and can
even become negative due to the non-monotonic behavior of the calculated
pore energy, while experiments typically report a single (positive) value
of the edge tension. This value could be viewed as the asymptotic edge tension obtained in the limit of large $r$, where the pore energy increases linearly with
$r$ and the edge tension is therefore constant.
Moreover, we find that the edge tension varies only little
with $\theta$ [see Figs.~\ref{FigLineTensionC}(a,c)]. This suggests that a non-zero $\theta$ has the primary effect of shifting up the curve for the pore energy, and only marginally distorts the variation of $G_p^{(c)}$ with $r$, which is also apparent from Figs.~\ref{FigPoreContr} and~\ref{FigPoreEcone}. However, changing the (relative) values
of $m$ and $h$ does have a pronounced effect on the numerical
values of the pore energy as well as on the edge tension [see Figs.~\ref{FigLineTensionC}(a,d)].

\subsection{Polygonal pores}
\label{secPolygPore}

\begin{figure}
\center
\includegraphics[width=8.2cm]{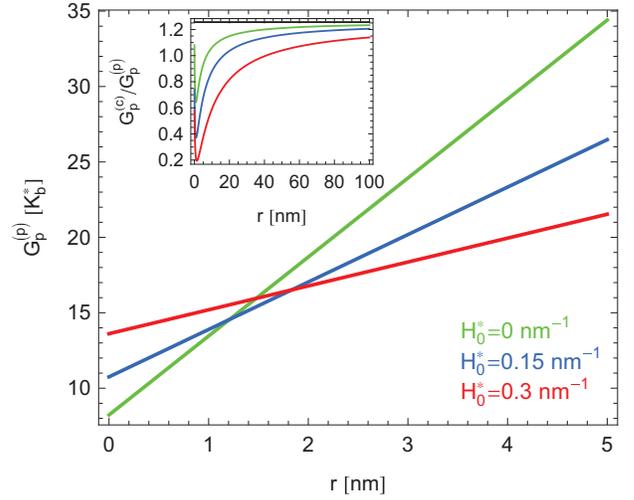}
\caption{\label{FigComparePoreE}(Color online) Elastic bending energy of a polygonal pore in Eq.~(\ref{polyPore}) versus pore radius $r$ for the icosahedron, and ratio of conical and polygonal pore energies (inset), for \cite{dubois04} $m=2$~nm and $h=0.5$~nm, using $H_0^\star=0$~nm$^{-1}$ for the green curves
(upper curves in the large-$r$ regime), $H_0^\star=0.15$~nm$^{-1}$ for the blue curves (middle curves in the large-$r$ regime), and $H_0^\star=0.3$~nm$^{-1}$ for the red curves (lower curves in the large-$r$ regime). The polygonal pore
energy is calculated by noting that, for the icosahedron, five ridges meet
at each vertex with the face angle $\beta_j=\pi/3$. The corresponding conical
pore energy is obtained using the vertex angle $\Omega=2\pi-5\arcsin(2/3)\approx2.6$
for the icosahedron, which gives $\theta \approx 0.2 \pi$. The horizontal
black line
in the inset denotes the ratio of the circumference of conical and polygonal
pores, which is equal to $\pi /\left(5 \sin \frac{\pi}{6}\right) \approx
1.3$
for the icosahedron.}
\end{figure}

Figure~\ref{FigComparePoreE} shows plots of the polygonal pore energy $G_p^{(p)}$
in Eq.~(\ref{polyPore}) as a function of the pore radius $r$ for the
vertex geometry of the icosahedron. A notable discrepancy between the polygonal pore energy
and the conical pore energy in Eq.~(\ref{GpFcal}) is that $G_p^{(p)}$ always
increases linearly with $r$ and, hence, does not lead to
an optimal pore radius for which the bending energy takes a minimal value.
However, Eq.~(\ref{polyPore}) gives a similar range of the pore energy as the corresponding expression for the bending energy of a conical pore. In particular, the asymptotic value of the
ratio $G_p^{(c)}/G_p^{(p)}$ is equal to the ratio of the pore circumferences in the two models, as indicated in the inset of Fig.~\ref{FigComparePoreE}. Finally, we note from Fig.~\ref{FigComparePoreE}
and Eqs.~(\ref{GP1cont}) and~(\ref{GP2cont}) that for small pore radii $r
\lessapprox 2$~nm
the contribution
$G_p^{(1)}$ to the polygonal pore energy dominates over the contribution
$G_p^{(2)}$, and vice versa. As a result, for
the parameter values in Fig.~\ref{FigComparePoreE}, the polygonal pore energy
increases
with increasing spontaneous curvature for small pore radii, but decreases with increasing spontaneous curvature for large pore radii.

\section{Polyhedral bending energies}
\label{secPolyE}

In this section we evaluate the total bending energy of bilayer polyhedra as a function of the polyhedron radius~$R_p$. As in Secs.~\ref{secBendE} and~\ref{secAsym}, the polyhedron radius is defined \cite{lidmar03} through $A=4 \pi R_p^2$, where $A$ is the polyhedron area, which is in turn proportional to the polyhedron
ridge length with a proportionality constant characteristic of the polyhedral geometry.
The total bending energies associated with different
polyhedral shapes are compared for a fixed area rather than a fixed volume since, as discussed from a theoretical perspective in Sec.~\ref{secVtable} and also observed in experiments
\cite{glinel04,dubois04,dubois01}, closed bilayer vertices are expected to break up to form pores, thus allowing adjustment of the polyhedron volume for a given number of amphiphiles or fixed surface area. Following Secs.~\ref{secBendE} and~\ref{secAsym}, polyhedral bending energies involve contributions due to ridges and vertices.
The vertex part of the polyhedron bending energy is independent of
the polyhedron size, and will generally favor bilayer polyhedra which only involve a few vertices. The ridge part of the polyhedron bending energy, however, increases with the polyhedron ridge length and, hence, with the
polyhedron radius.

Since ridges impose an energetic cost one expects that, for a fixed
area and dihedral angle, the faces of bilayer polyhedra relax to form regular polygons. While there are infinitely many regular convex polygons, there are only five regular convex polyhedra---the Platonic solids, which are vertex-transitive, edge-transitive, and face-transitive \cite{cromwell97,coxeter80,weisstein09}. Thus, all vertices, ridges, and faces of any given Platonic solid share the same geometric properties relating, for instance, to the values of face and dihedral angles. A natural generalization of the Platonic solids are the semiregular polyhedra, which are vertex-transitive and have regular (but not necessarily congruent) polygons as faces. Apart from the Platonic solids, the semiregular convex polyhedra encompass the 13 Archimedean solids and the two (infinitely large) families of prisms and antiprisms. Relaxing the constraint of vertex-transitivity, one obtains the class of convex polyhedra with regular
polygons as faces. In addition to the Platonic solids, Archimedean solids,
prisms, and antiprisms, this class includes the 92 Johnson solids. It has
been shown \cite{berman71,proofJZ} that this list exhausts all convex polyhedra with regular faces.

\begin{figure}[!]
\center
\includegraphics[width=7.87cm]{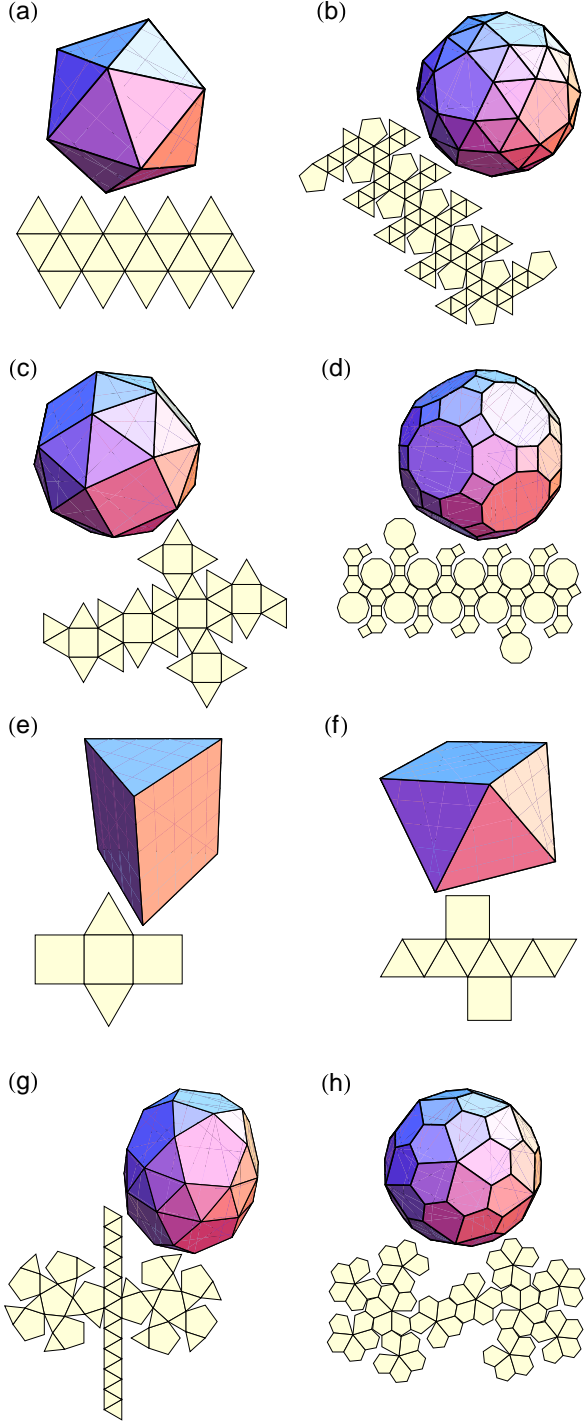}
\caption{\label{figIllPoly}(Color online) Image and net representations of (a) the icosahedron, (b) the snub
dodecahedron, (c) the snub cube, (d) the great rhombicosidodecahedron, (e)
the triangular prism, (f) the square antiprism, (g) the gyroelongated pentagonal
birotunda, and (h) the pentagonal hexecontahedron. Polyhedron (a) is a Platonic solid, polyhedra (b), (c), and (d) are Archimedean solids, polyhedron (e)
is a prism, polyhedron (f) an antiprism, polyhedron (g) a Johnson solid,
and polyhedron (h) a Catalan solid. The Catalan solid in (h) is the dual
of the Archimedean solid in (b), and the polyhedra in (b), (c), (g), and (h) are chiral.}
\end{figure}

Thus, counting prisms and antiprisms as one solid each, there are exactly 112 convex polyhedra with regular polygons as faces, and we will focus here on this set of polyhedra. As representative examples
of convex polyhedra with non-regular polygons as faces we will, however, also consider the bending energies of the Catalan solids, which are the duals of the Archimedean solids and, as such, are also highly symmetric. Figure~\ref{figIllPoly} shows examples of polyhedra belonging to the different symmetry classes
\cite{cromwell97,coxeter80,weisstein09,mathematica} we are concerned
with here. In particular, Sec.~\ref{secHomE} presents results pertaining to the bending energy of homogeneous bilayer polyhedra, and Sec.~\ref{secHetE} focuses
on the bending energy of bilayer polyhedra exhibiting segregation of excess amphiphiles. Finally, Sec.~\ref{secGen} discusses to what extent our results regarding the minimal bending energies of bilayer polyhedra can be expected to be valid for ridge, vertex, and pore energies which deviate from the elastic models developed in Secs.~\ref{secBendE} and~\ref{secAsym}.

\subsection{Homogeneous polyhedra}
\label{secHomE}

\begin{figure}[!]
\center
\includegraphics[width=8.5cm]{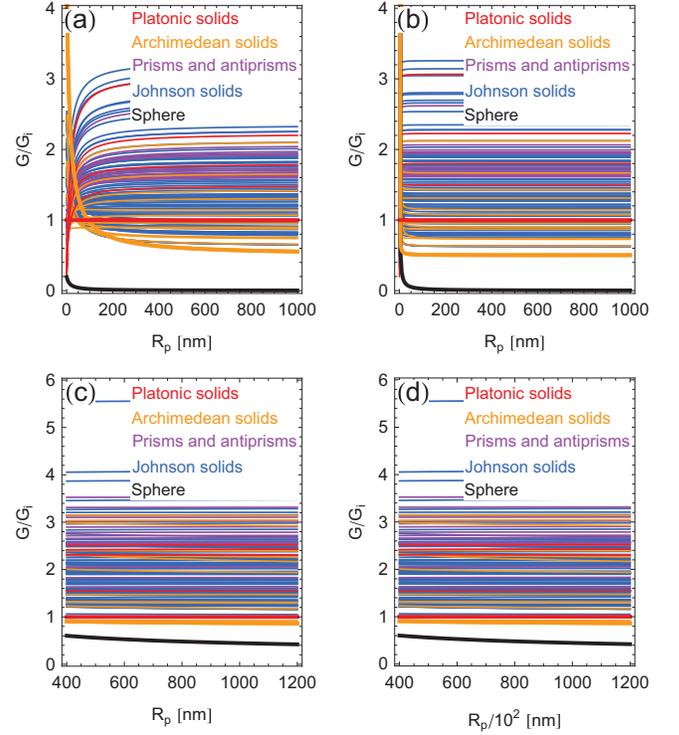}
\caption{\label{HomReg}(Color online) Total bending energies of the convex polyhedra
with regular faces, normalized by the bending energy of the icosahedron,
$G_i$, for homogeneous bilayers with (a) the vertex energy $G_v^{(h)}$ in Eq.~(\ref{EVgen}) and the ridge energy $G_r^{(h)}$ in Eq.~(\ref{ridgeE}), (b) the polygonal
pore energy $G_p^{(p)}$ in Eq.~(\ref{polyPore}) with $r=0$ and the ridge energy $G_r^{(h)}$ in Eq.~(\ref{ridgeE}), (c) the polygonal
pore energy $G_p^{(p)}$ in Eq.~(\ref{polyPore}) with $r=0$ and the upper bound
$Y/K_b=10$~nm$^{-2}$
on the ridge energy $G_r^{(LW)}$ in Eq.~(\ref{ridgeELW}), and (d) the polygonal
pore energy $G_p^{(p)}$ in Eq.~(\ref{polyPore}) with $r=0$ and the lower bound
$Y/K_b=10^{-3}$~nm$^{-2}$
on the ridge energy $G_r^{(LW)}$ in Eq.~(\ref{ridgeELW}). We use the parameter values \cite{dubois01,dubois04,hartmann06} $m=2$~nm, $h=0.5$~nm, $K_b^\star=K_b/100$, and $H_0^\star=0$~nm$^{-1}$. The bold black curve denotes the bending energy of the sphere, and the colored (gray) curves denote the bending energies of bilayer polyhedra, where the bold curve minimizing polyhedral bending energy in
the large-$R_p$ regime corresponds to the snub dodecahedron.}
\end{figure}

Figure~\ref{HomReg} shows the bending energies of the convex polyhedra with regular faces as a function of the polyhedron radius $R_p$. To begin, consider
homogeneous polyhedra with closed bilayer vertices [see Fig.~\ref{HomReg}(a)].
We calculate the relevant bending energies using the expression $G_r^{(h)}$ in Eq.~(\ref{ridgeE}) for the ridge energy and $G_v^{(h)}$ in Eq.~(\ref{EVgen}) for the vertex energy, together with the geometric parameters characterizing
the convex polyhedra with regular polygons as faces 
\cite{cromwell97,coxeter80,weisstein09,mathematica}. From Fig.~\ref{HomReg}(a) one finds that spherical bilayer vesicles have a lower bending energy than any of the polyhedral symmetries considered. Moreover, in agreement with a previous study \cite{dubois04}, we find
that the icosahedron [see Fig.~\ref{figIllPoly}(a)] minimizes bending energy among the Platonic solids.
However, as shown in Fig.~\ref{HomReg}(a), the icosahedron does not minimize bending energy if one allows for
more general polyhedral symmetries.

As noted above, closed bilayer vertices may break up to form (closed) pores, and the relevant energy curves are displayed in Fig.~\ref{HomReg}(b). These curves are again obtained with the
ridge energy $G_r^{(h)}$ in Eq.~(\ref{ridgeE}), but now this expression is
combined with
the polygonal pore energy $G_p^{(p)}$ in Eq.~(\ref{polyPore}) for $r=0$.
Although the details of the results in Fig.~\ref{HomReg}(b)
are quantitatively different from those in Fig.~\ref{HomReg}(a), we again find that in general the sphere is energetically favorable over the convex polyhedra
with regular faces, and that the icosahedron does not minimize elastic bending
energy among arbitrary polyhedral shapes. Increasing the pore radius does not
change these conclusions.

In Figs.~\ref{HomReg}(c,d) we plot polyhedral bending energy as a function
of $R_p$ using the ridge energy $G_r^{(LW)}$ in Eq.~(\ref{ridgeELW}) and
the pore energy $G_p^{(p)}$ in Eq.~(\ref{polyPore}) with $r=0$. As already mentioned
above, simulations suggest \cite{lidmar03,nguyen05}
that, at least for the icosahedron, $G_r^{(LW)}$ gives a good description
of the ridge energy for $\Gamma\gtrapprox 10^7$. Using the somewhat less stringent criterion $\Gamma > 10^6$, this then implies $R_p \gtrapprox 400$~nm for the upper bound $Y/K_b=10$~nm$^{-2}$ [see Fig.~\ref{HomReg}(c)]. We also
find with this modified expression of the ridge energy that spherical bilayer
vesicles have lower bending energy than any polyhedral shape considered, and that the icosahedron
does not represent the polyhedral shape with minimal bending energy. Applying the lower bound $Y/K_b=10^{-3}$~nm$^{-2}$
has the effect of shifting the curves for the ridge energy to larger polyhedron
radii, and does
not modify these conclusions. The corresponding results are shown in Fig.~\ref{HomReg}(d).

What is the polyhedral shape that minimizes the elastic bending energy among the convex polyhedra with
regular faces? As apparent from Fig.~\ref{HomReg}, the answer to this question
will generally depend on the polyhedron size and the particular expression
of the polyhedron
energy considered. Indeed, in the limit $R_p \to \infty$ the icosahedron only represents the 34th-lowest energy shape for the ridge energy $G_r^{(h)}$ in Eq.~(\ref{ridgeE}), but the third-lowest energy shape for $G_r^{(LW)}$ in Eq.~(\ref{ridgeELW}). However, for large enough polyhedron sizes, the snub dodecahedron [see Fig.~\ref{figIllPoly}(b)] minimizes polyhedral
bending energy for all ridge energies considered in Fig.~\ref{HomReg}.
Moreover, independently of the particular expression of the polyhedral bending energy
used, the snub cube [see Fig.~\ref{figIllPoly}(c)] also has a lower elastic bending energy than the icosahedron in this limit. For the scenarios considered
in panels (a), (b), and (c) of Fig.~\ref{HomReg}, this asymptotic behavior already manifests itself for the typical polyhedron size $R_p \approx 500$~nm observed in experiments \cite{dubois01,dubois04,glinel04,meister03,vautrin04},
while the lower bound $Y/K_b=10^{-3}$~nm$^{-2}$ in Fig.~\ref{HomReg}(d) only
applies to polyhedron sizes much larger than the observed size of bilayer
polyhedra.

\begin{figure}[!]
\center
\includegraphics[width=8.5cm]{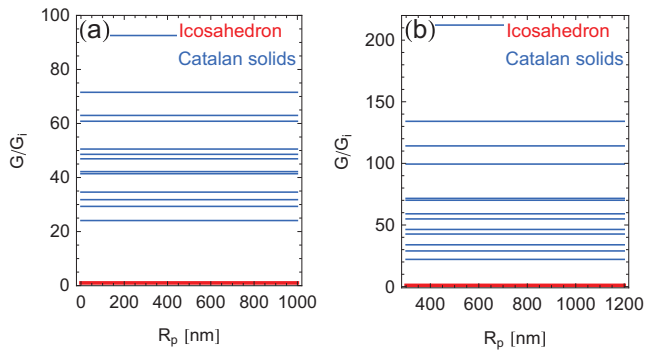}
\caption{\label{HomCat}(Color online) Ridge energies of the 13 Catalan solids, normalized by the ridge energy of the icosahedron, $G_i$, with (a) the ridge energy $G_r^{(h)}$ in Eq.~(\ref{ridgeE}), and (b) the upper bound $Y/K_b=10$~nm$^{-2}$ on the ridge energy $G_r^{(LW)}$ in Eq.~(\ref{ridgeELW}).}
\end{figure}

In Fig.~\ref{HomCat} we compare the total ridge energies of the 13 Catalan solids to the total ridge energy of the icosahedron as a function of the polyhedron radius. Panel (a) of Fig.~\ref{HomCat}
is obtained using the ridge energy $G_r^{(h)}$ in Eq.~(\ref{ridgeE}), whereas panel (b) corresponds to the ridge energy $G_r^{(LW)}$ in Eq.~(\ref{ridgeELW}) with the upper or, analogously, the lower bound on $Y/K_b$. Comparison
of Fig.~\ref{HomCat} with the large-$R_p$ regime in Fig.~\ref{HomReg}
shows that, as already anticipated on intuitive grounds, the total ridge energies of the Catalan solids are indeed much larger than those of the convex polyhedra with regular polygons as faces.

\subsection{Heterogeneous polyhedra}
\label{secHetE}

Perhaps the most basic result of the above analysis of the elastic bending
energies of homogeneous bilayer vesicles is that, independently of the particular expression of the polyhedral bending energy considered, spherical bilayer vesicles allow (much) lower bending energies than bilayer polyhedra. However, according to the experimental
observations in Refs.~\cite{dubois04,dubois01,glinel04}, pores are seeded into bilayers via molecular segregation if there is a slight excess of one amphiphile
species over the other. Thus, pores can have a role beyond reducing the elastic bending energy associated with the vertices of bilayer polyhedra.
How do the total bending energies of bilayer
polyhedra compare to the bending energies of spherical bilayer vesicles having an equal (or greater) number of pores? This question is addressed most conveniently by eliminating the vertex energies
altogether, and only comparing polyhedral ridge energies to the total
elastic energy associated with spherical bilayer vesicles.

\begin{figure}[!]
\center
\includegraphics[width=8.5cm]{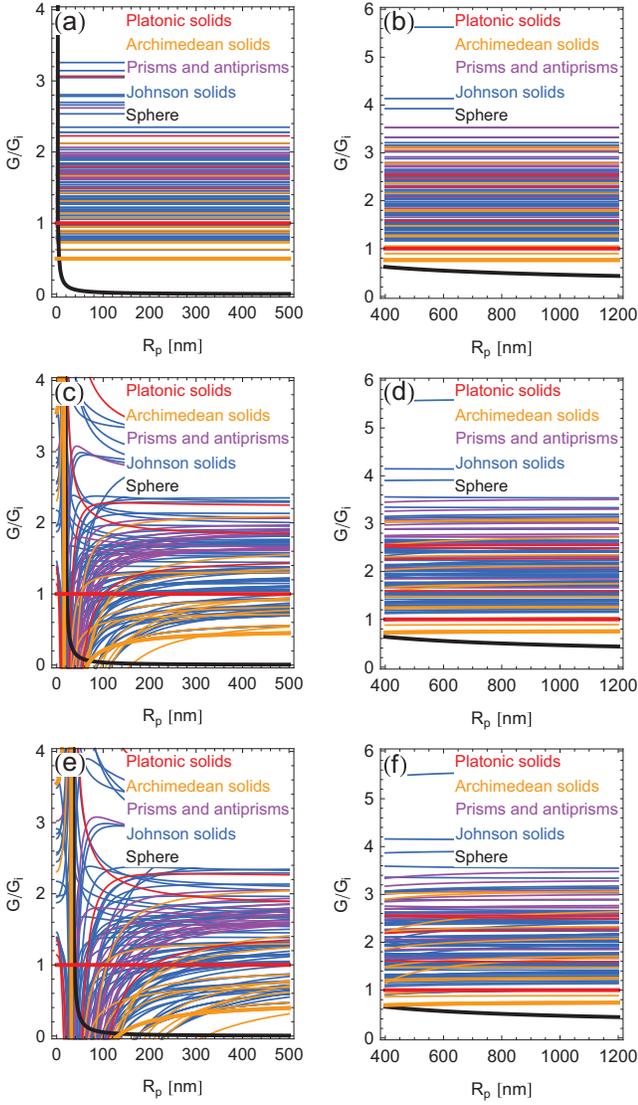}
\caption{\label{Het1}(Color online) Total bending energies of the convex polyhedra
with regular faces, normalized by the bending energy of the icosahedron,
$G_i$, with segregation of excess amphiphiles at vertices but not at ridges.
The energy curves are obtained with the ridge energy $G_r^{(h)}$ in Eq.~(\ref{ridgeE})
[panels (a), (c), and (e)] and the upper bound $Y/K_b=10$~nm$^{-2}$ on the ridge energy $G_r^{(LW)}$ in Eq.~(\ref{ridgeELW}) [panels (b), (d), and (f)]
with pores of radius (a,b) $r=0$, (c,d) $r=20$~nm, and (e,f) $r=40$~nm at
each vertex. The bold black curve denotes the bending energy of the sphere, and the colored (gray) curves denote the bending energies of bilayer polyhedra, where the bold curve minimizing polyhedral bending energy in
the large-$R_p$ regime corresponds to the snub dodecahedron.}
\end{figure}

Allowing molecular segregation at pores only, and setting the pore radius
equal to zero, we again find that the sphere is energetically favorable over any polyhedral shape
for physically relevant values of the polyhedron radius [see Figs.~\ref{Het1}(a,b)]. This conclusion holds
for the ridge energy $G_r^{(h)}$ in Eq.~(\ref{ridgeE}) [see Fig.~\ref{Het1}(a)] as well as for the ridge energy $G_r^{(LW)}$ in Eq.~(\ref{ridgeELW}) [see Fig.~\ref{Het1}(b)]. Furthermore, Figs.~\ref{Het1}(a,b) imply that, even if there is no energetic cost
associated with the vertices of polyhedral bilayer vesicles, spherical bilayer vesicles
are still energetically favorable.  The latter point
is particularly relevant considering that, in analogy to pores forming
in planar membranes~\cite{jackson09}, the conical and polygonal pore geometries we have considered here
may not represent general minima of polyhedral pore energies.

Figures~\ref{Het1}(c,d) and~\ref{Het1}(e,f) show the bending energy of bilayer
polyhedra with molecular segregation at pores of radius $r=20$~nm and $r=40$~nm, respectively, using the ridge energies $G_r^{(h)}$ in Eq.~(\ref{ridgeE})
and $G_r^{(LW)}$ in Eq.~(\ref{ridgeELW}). The plots in Figs.~\ref{Het1}(c,d) thereby
correspond to the typical polyhedra ($R_p \approx 500$~nm) and pore ($r \approx 20$~nm) sizes reported in Refs.~\cite{dubois01,dubois04,glinel04,meister03,vautrin04}. We note that the polyhedral ridge length decreases with increasing~$r$,
leading to a reduction in the polyhedral ridge energy. Hence, we expect that the total polyhedral ridge energy decreases with increasing $r$ and, indeed, approaches
zero as $2 r$ approaches the ridge length. This is borne
out by the results in Fig.~\ref{Het1}. However, the results in
Fig.~\ref{Het1} also suggest that, for bilayer polyhedra which only exhibit molecular segregation at vertices, the regime for which polyhedral bending energies are smaller than the bending energy of the sphere is, at
best, very narrow. Thus, molecular segregation of excess amphiphiles at polyhedral vertices is not expected to be sufficient to stabilize
polyhedral bilayer vesicles over spherical bilayer vesicles, even in the somewhat artificial
limit of molecular segregation at pores which are very large in relation to the total polyhedron size.

\begin{figure}[!]
\center
\includegraphics[width=8.48cm]{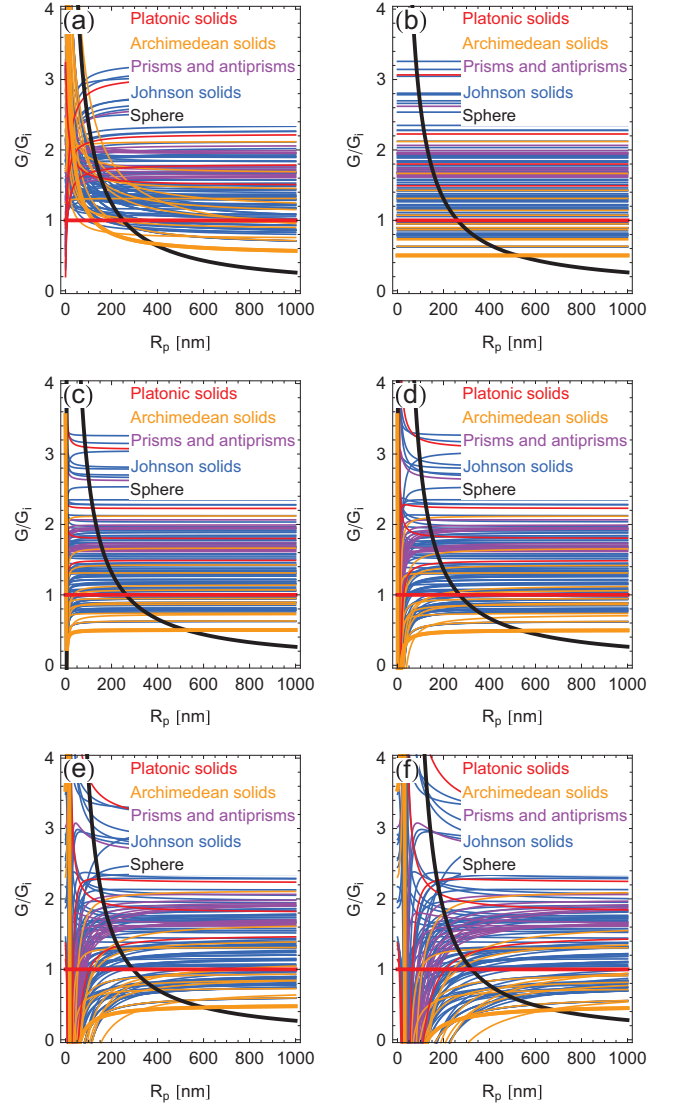}
\caption{\label{Het2}(Color online) Total bending energies of the convex polyhedra with regular faces, normalized by the bending energy of the icosahedron, $G_i$, with segregation of excess amphiphiles at ridges but not pores [panel (a)], and at ridges and pores [panels (b)--(f)].
For panel (a) we use the pore energy $G_p^{(p)}$ in Eq.~(\ref{polyPore}) with $r=0$ and the ridge energy $G_r^{(s)}$ in Eq.~(\ref{modRE}) with the parameter
values \cite{dubois01,dubois04,hartmann06} $m=2$~nm, $h=0.5$~nm, $K_b^\star=K_b/100$, and $H_0^\star=0$~nm$^{-1}$. The remaining panels are obtained using only the ridge
energy $G_r^{(s)}$ in Eq.~(\ref{modRE}) with pores of radius (b) $r=0$,
(c) $r=1$~nm, (d) $r=5$~nm, (e) $r=20$~nm, and (f) $r=40$~nm. The bold black curve denotes the bending energy of the sphere, and the colored (gray) curves denote the bending energies of bilayer polyhedra, where the bold curve minimizing polyhedral bending energy in
the large-$R_p$ regime corresponds to the snub dodecahedron.}
\end{figure}

Figure~\ref{Het2} shows the elastic bending energies of the convex polyhedra
with regular faces obtained
with the ridge energy $G_r^{(s)}$ in Eq.~(\ref{modRE}) for perfect molecular segregation along ridges. First, consider the case in which there
is molecular segregation at ridges, but not at pores [see Fig.~\ref{Het2}(a)].
Using the pore energy $G_p^{(p)}$ in Eq.~(\ref{polyPore}) with $r=0$, we find a pronounced regime for which polyhedral bilayer vesicles are energetically favorable over spherical bilayer vesicles ($R_p\lessapprox 400$~nm). For small $R_p$ there is a narrow regime for which the icosahedron is the polyhedral shape with minimal bending energy, while there are more prominent regimes at larger polyhedron sizes for which the snub cube and the snub dodecahedron are the polyhedral shapes minimizing bending energy. Thus,
molecular segregation along \textit{ridges} is found to be crucial for the stabilization of polyhedral bilayer vesicles over spherical bilayer vesicles.

Allowing molecular segregation at pores as well as ridges, one obtains
\cite{footnotePRL} the bending energies shown in Figs.~\ref{Het2}(b--f).
We find a pronounced regime $R_p\lessapprox600$~nm for which polyhedral bilayer
vesicles are energetically favorable compared to spherical bilayer vesicles if the same number of
pores is seeded into all vesicles. The polyhedral shape which generally minimizes
elastic bending energy for the typical polyhedron size $R_p \approx 500$~nm and pore
size $r \approx 20$~nm observed
in experiments \cite{dubois01,dubois04,glinel04,meister03,vautrin04} is the snub dodecahedron. Moreover, for large pore radii a sequence of polyhedral shapes with minimal
bending energy is obtained as a function of pore radius.
The most notable of these polyhedral shapes is the great rhombicosidodecahedron
[see Fig.~\ref{figIllPoly}(c)], which surpasses the snub dodecahedron in bending energy at $R_p \approx 300$~nm [Fig.~\ref{Het2}(e)] or at $R_p \approx 600$~nm [Fig.~\ref{Het2}(f)].

\begin{figure}[!]
\center
\includegraphics[width=8.5cm]{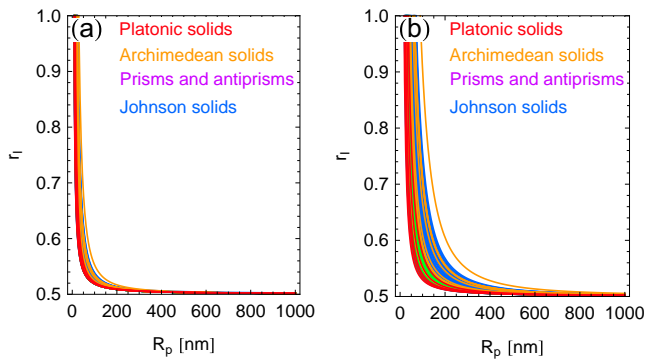}
 \caption{\label{figImbal}(Color online) Theoretical estimates of the optimal amphiphile imbalance $r_I$,
defined in Eq.~(\ref{estrI}), for the convex polyhedra with regular faces as a function of the polyhedron radius $R_p$
with \cite{dubois04} $m=2$~nm for (a) $r=0$~nm and (b) $r=20$~nm.}
\end{figure}

As discussed in Sec.~\ref{secSeg}, the model of perfect molecular segregation used for Figs.~\ref{Het1} and~\ref{Het2} allows us to obtain a phenomenological estimate
of the optimal
amount of excess amphiphiles for a given polyhedral shape and size.
Figure~\ref{figImbal} shows plots of the ratio of the amphiphile species in excess to the total amphiphile
content as a function of the polyhedron
radius $R_p$ for the convex polyhedra with regular polygons as faces. For the typical polyhedron radius $R_p \approx 500$~nm and pore radius $r \approx 20$~nm observed in
experiments \cite{dubois01,dubois04,glinel04,meister03,vautrin04} we find $r_I \approx 0.51$ as the optimal imbalance in the concentrations of the two amphiphile
species. The corresponding experimental estimate
is $r_I \approx 0.57$
\cite{dubois01,dubois04}. We expect that in experiments not all excess amphiphiles are segregated
along the ridges and vertices of polyhedra as a result of, for instance, entropic mixing within
bilayer polyhedra or the formation of micelles~\cite{dubois04}. Thus, our theoretical estimate of $r_I$ is in reasonable accord with the experimental results given the level of approximation
involved in making such estimates.

\subsection{Generalized ridge energy}
\label{secGen}

In Secs.~\ref{secHomE} and~\ref{secHetE} we found that, for large enough
polyhedron sizes, the snub dodecahedron minimizes bending energy among the
convex polyhedra with regular polygons as faces. This result was obtained
with the heuristic expressions of the ridge energy in Eqs.~(\ref{ridgeE})
and~(\ref{modRE}), and also with the limiting expression of the ridge energy in Eq.~(\ref{ridgeELW}). Does this conclusion regarding the polyhedral shape with minimal bending energy also hold for more general expressions
of the ridge energy? To address this question, consider ridge energies of the form
\begin{equation}
G_r \propto (\pi-\alpha_i)^p l_i^q\,,
\end{equation}
with $p=2$ and $q=1$ corresponding to
Eqs.~(\ref{ridgeE}) and~(\ref{modRE}),
and $p=7/3$ and $q=1/3$ corresponding to Eq.~(\ref{ridgeELW}). As before, we seek the minimum energy shape among the convex polyhedra with
regular faces, but now as a function of $p$ and $q$. Numerically one finds that, if $q$ is chosen small ($q\lessapprox 0.2$) or large ($q\gtrapprox 1.6$) enough, one can have $p\gtrapprox2$ with the snub dodecahedron no longer being the minimum energy shape.

What physical scenarios could lead to a ridge energy with values of $p$ and
$q$ so that the snub dodecahedron does not correspond to the most favorable
polyhedral shape? As discussed in Secs.~\ref{secBendE} and \ref{secAsym},
the available expressions of the ridge energy obtained from elasticity theory
firmly lie within the regime for which the snub dodecahedron minimizes elastic bending energy for large polyhedron sizes. However, segregation of excess amphiphiles implies that bilayer polyhedra are locally charged. Hence, electrostatic interactions could, in
principle, affect the symmetry of bilayer polyhedra \cite{dubois01,dubois04} and modify the ridge energy \cite{betterton99,mbamala06,mengistu08}. In Appendix~\ref{appES} we provide a simple
example of how electrostatic interactions could lead to an expression of
the ridge energy which is qualitatively different
from the elastic ridge energies considered in Secs.~\ref{secBendE} and~\ref{secAsym}, resulting in a polyhedron other than the snub dodecahedron as the energetically most favorable polyhedral shape for large vesicle sizes.

\section{Discussion}
\label{secDisc}

In agreement with expectations based on the classic framework for describing and predicting vesicle shape \cite{boal02,seifert97,safran03,phillips09}, our calculations imply that vesicles with smooth curvature are favorable over polyhedral vesicles for bilayers of uniform composition. However, allowing for molecular segregation of excess amphiphiles with high spontaneous curvature we find, consistent with the experimental phenomenology of bilayer polyhedra \cite{dubois01,dubois04,glinel04,meister03,vautrin04}, that polyhedral bilayer vesicles can have lower elastic bending energy than spherical bilayer vesicles. Furthermore, on the basis of our calculations we expect bilayer vertices to be unstable to the formation of (closed) pores. Again, this result is in agreement with experimental observations \cite{dubois01,dubois04,glinel04}, and suggests that bilayer polyhedra are
permeable.

According to our theoretical analysis, the mechanism lowering the bending energy of polyhedral bilayer vesicles below the bending energy of spherical
bilayer vesicles is segregation of excess amphiphiles along the ridges of bilayer polyhedra as observed in Ref.~\cite{dubois04}. Segregation at pores, which was originally suggested in Ref.~\cite{dubois01} as a potential mechanism stabilizing polyhedral vesicle shapes, is not sufficient to produce polyhedra with bending energies which are favorable compared to the sphere for the
typical size of bilayer polyhedra observed in experiments 
\cite{dubois01,dubois04,glinel04,meister03,vautrin04}. Moreover, independent of the particular expressions of ridge, vertex, and pore energies used, we find that the icosahedron does not minimize bending energy among arbitrary polyhedral shapes. In fact, for large enough polyhedron sizes, the snub dodecahedron is the polyhedral shape minimizing bending energy among the convex polyhedra with regular faces, and the snub cube also has a lower bending energy than the icosahedron in this limit. This result can be understood on a qualitative level as arising from a trade-off between reduction of the total ridge energy via a decrease in the total ridge length, and an accompanying decrease in the dihedral angles associated with ridges, which in turn leads to an increase in the density of the ridge energy.

What sets the characteristic range of polyhedron sizes observed in experiments
\cite{dubois01,dubois04,glinel04,meister03,vautrin04}? As noted above, molecular segregation along ridges is crucial for the stabilization
of bilayer polyhedra. For molecular segregation to significantly lower the elastic bending energy of bilayer polyhedra,
there must be a sufficient number of excess amphiphiles to line polyhedral ridges \cite{footnoteDisc}. However, while the
polyhedral ridge length increases linearly with the polyhedron radius $R_p$, the number of excess amphiphiles increases
quadratically with $R_p$. In contrast, the bending energy of spherical bilayer vesicles is, to a first approximation, independent of $R_p$.
Thus, we speculate that the characteristic range of polyhedron sizes observed
in experiments roughly corresponds to the maximum polyhedron size which still gives a lower total
bending energy than the sphere. Following this simple heuristic argument, one obtains from Fig.~\ref{Het2} the characteristic polyhedron
size $R_p \approx 400$--$600$~nm, which lies at the lower end of the range
of polyhedron sizes reported in Refs.~\cite{dubois01,dubois04,glinel04,meister03,vautrin04}.

Our comparisons between the total elastic bending energies of polyhedral
and spherical bilayer vesicles relied crucially on the ridge energies in
Eqs.~(\ref{ridgeE}),~(\ref{modRE}), and~(\ref{ridgeELW}), respectively. The first two of
these expressions involve the parameter $d$ corresponding to the arc length
suspended by a ridge. We fixed this parameter, and an analogous parameter
appearing in the vertex energy in Eq.~(\ref{EVgen}), by assuming that ridges bend
over a spatial scale corresponding to only two inter-amphiphile spacings. Such molecularly sharp ridges are consistent with a polyhedral vesicle shape.
Also, with this choice of $d$, Eqs.~(\ref{ridgeE}) and (\ref{modRE}) are in
broad agreement with the ridge energy in Eq.~(\ref{ridgeELW}) obtained 
\cite{lobkovsky96,lobkovsky97,lidmar03,nguyen05}
for
a diverging F\"oppl-von K\'arm\'an number. However, one might question the validity of the Helfrich-Canham-Evans
free energy of bending for ridge and vertex geometries exhibiting large local curvature. Atomistic simulations \cite{hartmann06,noguchi03,wohlert06} would
potentially allow the systematic investigation of the limitations of the
simple continuum models of polyhedral ridges and vertices used here.

A more gradual bending of the amphiphile bilayer along ridges than assumed in Eqs.~(\ref{ridgeE}),~(\ref{modRE}), and~(\ref{ridgeELW})
would reduce the density of ridge energies. This, in turn, could potentially stabilize
facetted vesicles for polyhedron sizes larger than the maximum polyhedron
radii implied by our analysis. Experimental results
obtained on the basis of electron and light microscopy indeed suggest 
\cite{dubois01,dubois04,glinel04,meister03,vautrin04}
that larger sizes of facetted vesicles may be stable, and that these vesicles
exhibit ridges and vertices which bend more gradually than in the case of
truly polyhedral vesicles. However, the quantitative description
of such facetted vesicles calls for interacting ridge and vertex geometries, which we did not consider
in our simple elastic models of polyhedral ridges and vertices. Moreover, a more comprehensive understanding of the characteristic range of polyhedron sizes will, among other things, necessitate a quantitative description of the formation of bilayer polyhedra from spherical bilayer vesicles \cite{dubois01,glinel04,meister03,dubois04,vautrin04}
during the cooling down process. Such a description of kinetic effects \cite{hartmann06,noguchi03} will also be necessary to predict the distribution of the symmetries and sizes of polyhedral bilayer vesicles, and may shed light on the mechanism
leading to the segregation of excess amphiphiles.

Our investigation of the elastic energy of polyhedral bilayer vesicles was motivated by the proposal \cite{dubois04} that the shape of bilayer polyhedra is governed by minimization of elastic bending energy. The resulting expressions of the total polyhedral bending energy are obtained from simple models based on continuum elasticity theory, and do not consider the details of molecular interactions between amphiphiles. An approach complementary to the one developed here would therefore account for specific molecular structures \cite{nagle00} of amphiphile bilayers. In particular, as far as the symmetry of polyhedral bilayer vesicles is concerned, an intriguing possibility is that optimal tilt angles between amphiphiles, which have been reported for a variety of lipid species \cite{small86}, might influence the packing of amphiphiles along polyhedral vertices and ridges and, hence, affect the preferred vertex and ridge geometries. While such a detailed molecular study of the structure of bilayer polyhedra is beyond the scope of the present article, we note that a molecular-level approach is expected to suggest models of molecular segregation superior to the simple models of perfect segregation of excess amphiphiles employed here (see, for instance, Fig.~\ref{FigRidgeESeg}), and permit a more realistic representation of the amphiphile species used in experimental investigations of bilayer polyhedra \cite{dubois04,dubois01,meister03,glinel04,vautrin04}.

It is instructive to compare the results presented here to recent theoretical studies carried
out in the contexts of two-dimensional
superconductors with vortices \cite{dodgson96}, viral capsids \cite{zandi04,bruinsma03},
and the buckling of ionic
shells \cite{vernizzi07}, which all employed approaches complementary to
ours. In agreement with our analysis, these studies suggest that the elastic energies of chiral shapes such as the snub dodecahedron and the snub cube can be favorable compared to the icosahedron \cite{bruinsma03,zandi04,dodgson96} and that, even if the icosahedral shape is imposed, the minimum energy structure may still be chiral \cite{vernizzi07}. In our analysis we followed the experimental phenomenology of bilayer polyhedra \cite{dubois04,dubois01,meister03,glinel04,vautrin04} and focused on contributions to the elastic bending energy captured by the mean curvature. Thus, we neglected other contributions to the free energy of bilayer polyhedra stemming, for instance, from the Gaussian curvature, electrostatic interactions, or entropy loss due to molecular segregation. These other contributions to the free energy, as well as kinetic effects \cite{hartmann06,noguchi03} and the detailed molecular structure of amphiphile bilayers \cite{nagle00,small86} at polyhedral vertices and ridges, could potentially modify the preferred vesicle shape and polyhedral symmetry.

\section{Summary and conclusions}
\label{secSum}

In this article we explored the total elastic bending energies of polyhedral bilayer vesicles \cite{dubois01,glinel04,meister03,dubois04,vautrin04}. Due
to current experimental uncertainties regarding the physical
properties of bilayer polyhedra, we did not attempt to make accurate estimates of the absolute values of the elastic bending energy of polyhedral bilayer
vesicles. Instead, we made general predictions pertaining to the most favorable polyhedral symmetries, and to the competition between polyhedral and spherical bilayer vesicles. Our results only rely on broad assumptions concerning the mechanical properties of bilayer polyhedra, and the applicability of the Helfrich-Canham-Evans
free energy of bending \cite{canham70,helfrich73,evans74} at the ridges and vertices of bilayer polyhedra. We assessed the validity of these phenomenological expressions of ridge and vertex energies by making comparisons to solutions of the two-dimensional equations of elasticity obtained previously \cite{lobkovsky96,lobkovsky97,seung88,lidmar03,nguyen05,didonna02}
for polyhedral ridges and vertices in certain limiting cases.

In agreement with experiments on polyhedral bilayer vesicles 
\cite{dubois01,dubois04,glinel04,meister03,vautrin04}, we
find that bilayer polyhedra can indeed be energetically favorable compared to spherical bilayer vesicles if one allows for molecular segregation of excess amphiphiles
along the ridges of bilayer polyhedra. Furthermore, our calculations suggest that closed bilayer vertices may break up to form pores, which is
also consistent with experimental observations \cite{dubois04,dubois01,glinel04}. However, our analysis implies that, contrary to what has been suggested on the basis of experiments
\cite{dubois04,dubois01}, the icosahedron does not represent the
polyhedral shape with minimal bending energy among arbitrary polyhedral shapes
and sizes. Using a variety of different expressions of polyhedral bending energy we find that, for large polyhedron
sizes, the snub dodecahedron and the snub
cube have lower total bending energies than the icosahedron. Our results
suggest revisiting
the symmetry of polyhedral bilayer vesicles, and the possible mechanisms
governing their formation, in greater experimental detail.

\begin{acknowledgements}

This work was supported by a Collaborative Innovation Award of the Howard
Hughes Medical Institute, and the National Institutes of Health through NIH Award number R01 GM084211 and the Director's Pioneer Award. We thank A. Agrawal,
M. D. Betterton, M. B. Jackson, W. S. Klug, R. W. Pastor, T. R. Powers, D. C. Rees, M. H. B. Stowell, D. P. Tieleman, T. S. Ursell, D. Van
Valen, and H. Yin for helpful comments.

\end{acknowledgements}

\appendix

\section{Evaluation of conical pore energy}
\label{AppPore}

Substituting Eqs.~(\ref{R1eq})--(\ref{dSeq}) into the Helfrich-Canham-Evans
free energy of bending in Eq.~(\ref{helfrich}) one obtains an integral over $\omega$
with the integrand composed of a sum of terms proportional to $\cos \left(|\omega|\pm
\theta \right)/R_2$, $\cos \left(|\omega|\pm \theta \right)$, and $\cos \left(|\omega|\pm \theta \right) R_2$. The integrals corresponding to the latter
two terms can be evaluated by elementary methods. To evaluate the terms with integrands of the form $\cos \left(|\omega|\pm \theta \right)/R_2$ we note that
\begin{equation}
-\frac{(m-h) \cos(|\omega|\pm\theta)}{R_2}=
\frac{\cos^2(|\omega|\pm\theta)}{\xi-\cos(|\omega|\pm\theta)}
\end{equation}
and complete the square in the numerator on the right hand side of the above
expression. We then use the results \cite{gradshteyn80}
\begin{eqnarray} 
\int \frac{dx}{\xi- \cos x}&=&\frac{2}{\left(\xi^2-1\right)^{1/2}} \arctan \frac{\left(\xi^2-1\right)^{1/2}
\tan \frac{x}{2}}{\xi-1} \,,\nonumber \\&&\label{gradsht1}
\end{eqnarray}
valid for $\xi^2>1$, and
\begin{equation} 
\int dx \frac{ \cos x}{\xi- \cos x}=- x+\xi \int \frac{dx}{\xi- \cos x}\,,
\end{equation}
valid for $\xi- \cos x \neq 0$, to arrive at Eq.~(\ref{GpFcal}).

\section{Electrostatic ridge energy}
\label{appES}

Betterton and Brenner \cite{betterton99} analyzed the effect of electrostatics on the stability of planar membranes of fixed area. Assuming that the surface charge density is constant, the total
charge contained in the screening cloud surrounding the membrane in solution
is also constant. However, the
volume of the screening cloud depends on the membrane geometry. In particular,
formation of pores increases the volume accessible to counterions, thus
leading to an increase in entropy compared to planar membranes. The gain
in free energy due to pore formation can be quantified \cite{betterton99}
by noting that, for
$r\ll\lambda_D$, where $r$ is the pore radius and $\lambda_D$ is the Debye
length, the screening cloud gains a volume $2 \pi  \lambda_D r^2$ through
pore formation. Similarly, for $r\gg\lambda_D$, the volume change is $\pi^2  \lambda_D^2 r$. Approximating the strength of the electrostatic field in the screening cloud
by $E=\sigma/\epsilon_0 D$, where $\sigma$ is the surface charge density, $\epsilon_0$ is the electric constant, and $D$ is the dielectric constant, one therefore
expects an energy decrease of
\begin{equation} \label{bbestimate}
U = 
\begin{cases} \frac{\pi \sigma^2}{\epsilon_0 D} \lambda_D r^2 & \mbox{for}
\quad r\ll\lambda_D\,, \vspace*{0.1cm} \\ \frac{\pi^2 \sigma^2}{2 \epsilon_0
D} \lambda_D^2 r & \mbox{for} \quad r\gg\lambda_D \, ,
\end{cases}
\end{equation}
in which we have set the energy density of the electrostatic field equal to $\frac{1}{2}
\epsilon_0 D E^2$.

The heuristic estimates in Eq.~(\ref{bbestimate}) are
confirmed \cite{betterton99} by solving the Debye-H\"uckel equation for a
pore in a charged membrane. The electrostatic contributions to the free energy in Eq.~(\ref{bbestimate}) are in competition with the energy penalty imposed by line tension along the pore edge. As shown in Sec.~\ref{secAnPore},
the elastic pore energy is approximately linear in the pore radius beyond $r\approx 2$~nm. Thus, in principle there could be 
a regime for which pores of a finite radius are stable due to the competition
between elastic and electrostatic contributions to the free energy, although
attaining such a regime would require \cite{betterton99} delicate adjustment of the various elastic
and electrostatic parameters.

\begin{figure}[t!]
\center
\includegraphics[width=8.5cm]{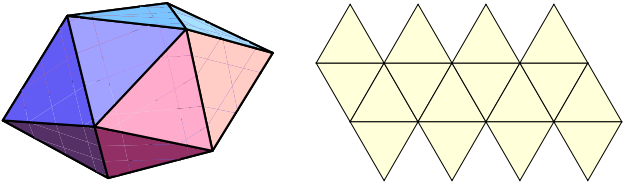}
\caption{\label{figpoly}(Color online) Image and net representations of the gyroelongated square dipyramid.}
\end{figure}

Based on the picture \cite{betterton99} outline above, we can obtain
a heuristic expression of the electrostatic ridge energy. Describing
a ridge as a bilayer bending by an angle $\pi-\alpha_i$ around a cylinder of
radius $R_1$ with charged amphiphiles in the outer membrane leaflet only
(see Fig.~\ref{FigRidgeESeg}), the volume of the screening cloud
associated with a ridge of length $l_i$ is approximately given by
\begin{equation}
\frac{\pi-\alpha_i}{2} \left(\lambda_D^2+2 R_1 \lambda_D \right) l_i\,,
\end{equation}
where we have followed the experimental observations in Refs.~\cite{dubois01,dubois04} and assumed that $\lambda_D \gg m$. Thus, one finds that the electrostatic energy is decreased by
\begin{eqnarray}
U=\frac{\sigma^2}{4 \epsilon_0 D} 
\sum_i \left(\lambda_D^2+2 R_1 \lambda_D \right) (\pi-\alpha_i) l_i \label{energyES}
\end{eqnarray}
through the formation of a ridge. In contrast to elastic ridge energies, the polyhedral shape with the most favorable electrostatic
ridge energy maximizes, for $R_1 \ll \lambda_D$, $\sum_i (\pi-\alpha_i) l_i$ within this heuristic picture.
Among the convex
polyhedra with regular polygons as faces, this is achieved by the gyroelongated square dipyramid shown in Fig.~\ref{figpoly}, while the snub dodecahedron
produces a somewhat smaller value of $\sum_i (\pi-\alpha_i) l_i$ than the icosahedron. However, based
on the experimental phenomenology of bilayer polyhedra,
electrostatic contributions to the free energy are expected \cite{dubois04}
to be negligible compared to elastic contributions. Furthermore, it is questionable
whether the assumption of a constant overall charge holds for bilayer polyhedra, and whether the mean-field picture invoked
here represents a good approximation of the energetics governing the narrow counterion clouds surrounding polyhedral ridges.


\begin{thebibliography}{99}
\bibitem{duboisTypo} The ridge energies in Eqs.~(2) and~(3) of Ref.~\cite{dubois04} appear to contain typos: In the notation of Ref.~\cite{dubois04}, $\alpha$ should be replaced by $\pi-\alpha$, and $k_c$ must be the bending modulus per unit length.
\bibitem{footnoteEr} These broad ranges are estimated on the basis of Refs.~\cite{dubois04,hartmann06}, which give the values $K_b \approx 10^3 k_B T$ and $Y \approx 0.4~$J/m$^2= 10^2 k_B T/$nm$^2$ for $1 k_B T=4\times10^{-21}$~J at $T=300K$.
\bibitem{footnoteDGerror} The corresponding plots of the pore energy for $\theta=0$ and resulting estimates of the optimal pore radius in Refs.~\cite{dubois04,glinel04}
do not seem to be correct.
\bibitem{footnoteDisc} For a large excess of one amphiphile species over the other, no closed vesicles are obtained \cite{dubois04}.
\end{thebibliography}
\end{document}